\documentclass[a4paper,11pt]{article}
\pdfoutput=1 

\usepackage[usenames,dvipsnames,svgnames,table]{xcolor} 
\usepackage{jcapmod}
\usepackage{verbatim}

\usepackage[T1]{fontenc} 
\usepackage[latin9]{inputenc}
\definecolor{lightgray}{gray}{0.9}
\definecolor{Amber}{rgb}{1.0, 0.75, 0.0}
\definecolor{blizzardblue}{rgb}{0.67, 0.9, 0.93}

\usepackage{float}
\usepackage{graphicx}
\usepackage{hyperref}
\usepackage{amsmath}
\usepackage{amssymb}
\usepackage{microtype}
\usepackage[shortlabels]{enumitem}
\usepackage{cancel}
\usepackage{amsbsy}
\usepackage{color}
\usepackage[capitalise]{cleveref}
\usepackage{breakurl}
\usepackage{mathtools}
\usepackage{csquotes}
\usepackage{lmodern}
\usepackage{bm}
\usepackage{dsfont}
\usepackage{bbold}
\usepackage{empheq}

\usepackage{caption}
\usepackage{subcaption}
\usepackage{placeins}
\usepackage{soul, xcolor}
\setstcolor{red}

\pdfpageheight\paperheight
\pdfpagewidth\paperwidth

\usepackage[textwidth=0.71\paperwidth, textheight=0.8\paperheight]{geometry}

\newcommand{\Kdelta}{\delta^{(\textrm{K})}}

\usepackage[table]{xcolor}

\allowdisplaybreaks

\makeatletter
\DeclareRobustCommand{\rcite}[1]{%
  \rcite@aux#1,\@nil{#1}%
}
\def\rcite@aux#1,#2\@nil#3{%
  \if\relax#2\relax
    Ref.~\cite{#3}%
  \else
    Refs.~\cite{#3}%
  \fi
}
\makeatother

\subheader{IFT-UAM/CSIC-24-47}

\title{\boldmath Cosmic topology. Part IVa. Classification of manifolds using machine learning: a case study with small toroidal universes}

\author[a]{Andrius Tamosiunas,}
\author[a]{Fernando Cornet-Gomez,}
\author[b,a,c]{Yashar Akrami,}
\author[d,e,f]{Stefano Anselmi,}
\author[b]{Javier Carr\'on Duque,}
\author[a]{Craig J. Copi,}
\author[g,c]{Johannes R. Eskilt,}
\author[a]{\"{O}zen\c{c} G\"{u}ng\"{o}r,}
\author[c]{Andrew H. Jaffe,}
\author[h]{Arthur Kosowsky,}
\author[b]{Mikel Martin Barandiaran,}
\author[a]{James B. Mertens,}
\author[a]{Deyan P. Mihaylov,}
\author[i]{Thiago S. Pereira,}
\author[a]{Samanta Saha,}
\author[a]{Amirhossein Samandar,}
\author[a]{Glenn D. Starkman,}
\author[a]{Quinn Taylor,}
\author[j]{and Valeri Vardanyan}

\collaboration{(COMPACT Collaboration)}
\collaborationImg{\includegraphics[width = 0.1\textwidth]{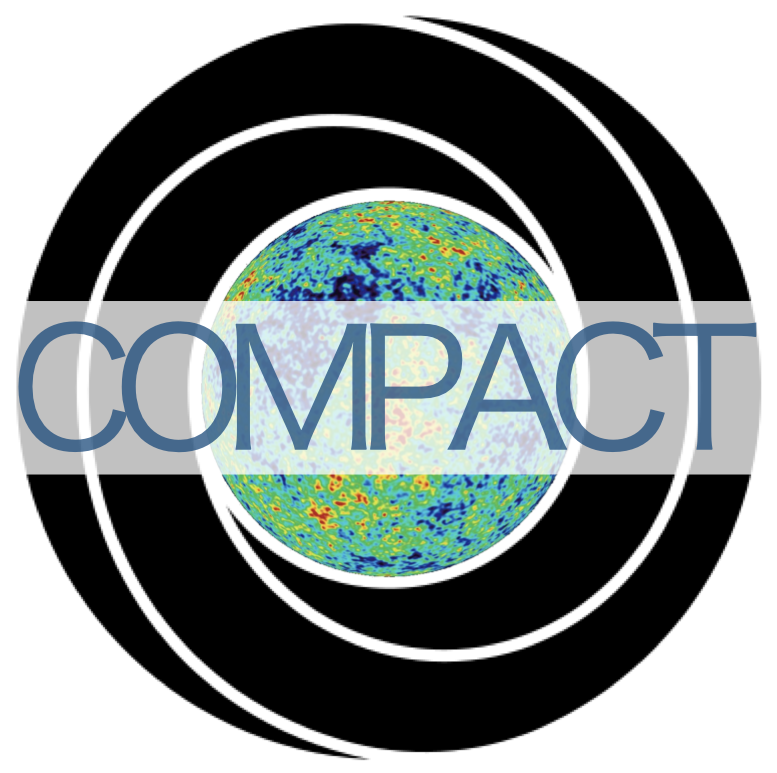}}

\affiliation[a]{CERCA/ISO, Department of Physics, Case Western Reserve University, 10900 Euclid Avenue, Cleveland, Ohio 44106, USA}
\affiliation[b]{Instituto de F\'isica Te\'orica (IFT) UAM-CSIC, C/ Nicol\'as Cabrera 13-15, Campus de Cantoblanco UAM, 28049 Madrid, Spain}
\affiliation[c]{Astrophysics Group \& Imperial Centre for Inference and Cosmology, Department of Physics, Imperial College London, Blackett Laboratory, Prince Consort Road, London SW7 2AZ, United Kingdom}
\affiliation[d]{INFN, Sezione di Padova, via Marzolo 8, I-35131 Padova, Italy}
\affiliation[e]{Dipartimento di Fisica e Astronomia ``G. Galilei'', Universit\`a degli Studi di Padova, via Marzolo 8, I-35131 Padova, Italy}
\affiliation[f]{Laboratoire Univers et Th\'eories, Observatoire de Paris, Universit\'e PSL, Universit\'e Paris Cit\'e, CNRS, F-92190 Meudon, France}
\affiliation[g]{Institute of Theoretical Astrophysics, University of Oslo, P.O. Box 1029 Blindern, N-0315 Oslo, Norway}
\affiliation[h]{Department of Physics and Astronomy, University of Pittsburgh, Pittsburgh, Pennsylvania 15260, USA}
\affiliation[i]{Departamento de F\'{i}sica, Universidade Estadual de Londrina,
Rod. Celso Garcia Cid, Km 380, 86057-970, Londrina, Paran\'a, Brazil}
\affiliation[j]{Kavli Institute for the Physics and Mathematics of the Universe (WPI), UTIAS, The University of Tokyo, Chiba 277-8583, Japan}

\emailAdd{andrius.tamosiunas@case.edu}
\emailAdd{fernando.cornetgomez@case.edu}
\emailAdd{yashar.akrami@csic.es}
\emailAdd{stefano.anselmi@pd.infn.it}
\emailAdd{javier.carron@csic.es}
\emailAdd{craig.copi@case.edu}
\emailAdd{j.r.eskilt@astro.uio.no}
\emailAdd{ozenc.gungor@case.edu}
\emailAdd{a.jaffe@imperial.ac.uk}
\emailAdd{kosowsky@pitt.edu}
\emailAdd{mikel.martin@uam.es}
\emailAdd{james.mertens@case.edu}
\emailAdd{deyan.mihaylov@case.edu}
\emailAdd{tspereira@uel.br}
\emailAdd{samanta.saha@case.edu}
\emailAdd{amirhossein.samandar@case.edu}
\emailAdd{glenn.starkman@case.edu}
\emailAdd{qxt42@case.edu}
\emailAdd{valeri.vardanyan@ipmu.jp}
\date{\today}

\abstract{
Non-trivial spatial topology of the Universe may give rise to potentially measurable signatures in the cosmic microwave background. 
We explore different machine learning approaches to classify harmonic-space realizations of the microwave background in the test case of Euclidean $E_1$ topology (the 3-torus) with a cubic fundamental domain of a size scale significantly smaller than the diameter of the last scattering surface. 
This is the first step toward developing a machine learning approach to classification of cosmic topology and likelihood-free inference of topological parameters.
Different machine learning approaches are capable of classifying the harmonic-space realizations with accuracy greater than 99\% 
if the topology scale is half of the diameter of the last-scattering surface and orientation of the topology is known.
For distinguishing random rotations of these sky realizations from realizations of the covering space, the extreme gradient boosting classifier algorithm performs best with an accuracy of 88\%. Slightly lower accuracies of 83\% to 87\% are obtained with the random forest classifier along with one- and two-dimensional convolutional neural networks.  The techniques presented here can also accurately classify non-rotated cubic $E_1$ topology realizations with a topology scale slightly larger than the diameter of the last-scattering surface, if enough training data are provided.
While information compressing methods like most machine learning approaches cannot
exceed the statistical power of a likelihood-based approach that captures
all available information, they potentially offer a computationally cheaper alternative.
A principle challenge appears to be accounting for arbitrary orientations of a given
topology, although this is also a significant hurdle for likelihood-based approaches.}


\begin{document}
\maketitle
\flushbottom

\section{Introduction}
\label{sec:introduction}

Einstein's general theory of relativity (GR) in combination with cosmological observations can be used to constrain the average local geometry of the Universe \cite{Einstein:1917ce, peebles:1993}.
A related but separate question is that of the global cosmic topology
(see, e.g., \rcite{Starkman:1998qx}).
A universe with non-trivial topology would possess a number of features generally not considered in the standard model of cosmology.
As an example, on a manifold with non-trivial topology, any point would have spacelike curves that could not be continuously deformed to a point \cite{Lachieze-Rey:1995, Luminet:1999}. 
If the length of the shortest such curve through the observer were short enough, this would result in an observer detecting a multitude of copies (\textit{clone images}) of astronomical sources \cite{Lehoucq:1996qe, Mota:2010jb, Fujii2011}.
For instance, light from a far-away galaxy would reach an observer via multiple paths, resulting in multiple observed images of the mentioned galaxy.
Similarly, if the length of the shortest closed loop through an observer were sufficiently less than the diameter of the last-scattering surface (LSS), an observer would detect the so-called circles-in-the-sky effect, referring to matched patterns in the cosmic microwave background (CMB) temperature fluctuations around pairs of circles on the celestial sphere \cite{Cornish:1996kv, Cornish:1997ab, Cornish:1997hz, Cornish:1997rp}.
Even if the topology scale were larger than the diameter of the LSS, there might still be information encoded in the CMB fluctuations, as discussed in \rcite{Riazuelo:2006tb,COMPACT:2022gbl,COMPACT:2022nsu, COMPACT:2023_paper2A}.
Several observational searches for these signatures of non-trivial topology have been performed, including Wilkinson Microwave Anisotropy Probe (WMAP) and \textit{Planck} searches for the circles-in-the-sky effect \cite{Cornish:2003db,ShapiroKey:2006hm, Vaudrevange:2012da, WMAP:2003elm, WMAP:2012nax, Planck:2013okc, Planck:2015gmu}.
Similarly, a Bayesian search that relies on comparing the pixel-pixel correlations in the observed CMB temperature map to those expected to be induced in manifolds with non-trivial topology was performed using \textit{Planck} survey data \cite{Souradeep1998arx, Planck:2013okc, Planck:2015gmu}. 
The outlined observational efforts, as of yet, have not detected any evidence for matched circle pairs in the CMB data. This, combined with the measurements of the local geometry of the Universe, allows us to constrain the set of allowed topology classes. Specifically, if the Universe is spatially flat, the set of allowed topologies consists of 18 classes (often denoted as $E_{1}$-$E_{17}$ for the Euclidean non-trivial topology classes and $E_{18}$ for the trivial topology class, i.e., the covering space). 
Each of the 18 classes can then have up to 6 free parameters, corresponding approximately to the lengths of the sides of the fundamental domain\footnote{
    Technically, we should refer to the Dirichlet domain, which is a specific fundamental domain. For details see \rcite{COMPACT:2023_paper2A}.
}
and the angles between them, plus 6 additional possible degrees of freedom corresponding to the position and orientation of the observer \cite{Lachieze-Rey1995pr, Luminet1999, hitchman2009geometry, Riazuelo2004:prd, thurston2014three}.
The key challenge from an observational perspective then is to distinguish the effects of these 18 topology classes on the CMB anisotropies.

The CMB temperature fluctuations $\Delta T$ can be expanded in terms of spherical harmonics, such that
\begin{equation}
\Delta T(\theta, \phi)=\sum_{\ell, m} a_{\ell m} Y_{\ell m}(\theta, \phi),
\label{eq:Spherical}
\end{equation}

\noindent where $\theta$ and $\phi$ are the polar and azimuthal angles on the sky, $Y_{\ell m}$ are the spherical harmonics, and $a_{\ell m}$ are the complex harmonic-space coefficients with $\ell$ and $m$ the multipole and azimuthal numbers, respectively.
Having decomposed the observed CMB temperature fluctuation pattern in spherical harmonics, we note that, if the sky is the result of a Gaussian process, then all the information about the temperature anisotropies is captured by the 2-point angular correlation matrix $C_{\ell m \ell^{\prime} m^{\prime}} = \langle a_{\ell m} a_{\ell^{\prime} m^{\prime}}^{*} \rangle$. 
In the case of trivial topology, which is isotropic, this correlation matrix is diagonal,  $C_{\ell m \ell^{\prime} m^{\prime}} = C_{\ell} \delta_{\ell \ell^{\prime}}^{(\mathrm{K})} \delta_{m m^{\prime}}^{(\mathrm{K})}$, where $C_{\ell}$ is the angular power spectrum and $\Kdelta_{ij}$ is the Kronecker delta.   
However, this is not the case when considering non-trivial topology -- the assumption of statistical isotropy is broken, resulting in non-zero off-diagonal terms in the correlation matrix. 

Given that these non-diagonal correlations are a key signature of non-trival topology, significant work has gone into understanding and classifying such features, e.g., in \rcite{Riazuelo2004:prd, COMPACT:2022gbl, COMPACT:2022nsu, COMPACT:2023_paper2A}. 
One can also quantify the information related to non-trivial topologies in the form of Kullback-Leibler (KL) divergence for given correlation matrices. 
Previous work \cite{COMPACT:2022gbl, COMPACT:2022nsu, COMPACT:2023_paper2A} has demonstrated that even for a non-trivial topology with a scale slightly larger than the diameter of the LSS, the KL divergence is larger than unity, i.e., a given CMB realization from a non-trivial topology could still, in principle, be distinguished from the one from the trivial topology. 
However,  even though we can demonstrate that the information related to non-trivial topologies exists and is encoded in the correlations between the different harmonic coefficients, deducing the topology classes based on an individual $a_{\ell m}$ realization is a complicated inverse problem. 
This problem cannot generally be approached analytically and other techniques, such as those based on artificial intelligence, are likely required.

In this work we present a set of machine learning algorithms to distinguish and classify CMB realizations from different topologies in harmonic-space ($a_{\ell m}$ coefficients) and correlation space ($a_{\ell m} a_{\ell^{\prime} m^{\prime}}^{*}$ for each realization). While the problem of distinguishing different topologies using harmonic-space realizations or CMB maps has been considered before (e.g., in \rcite{Kunz:2006, Aurich2024}), 
to the best of our knowledge this is the first application of machine learning  techniques to this problem. 
We start by numerically generating a set of harmonic-space realizations for the cubic $E_{1}$ topology  of different sizes, where by size we refer to the length of the fundamental domain in units of the diameter of the LSS. 
Specifically, we generate cubic $E_{1}$ realizations in four size classes, $ L \in \{ 0.05, 0.1, 0.5, \infty  \} \times L_{\rm LSS}$, where $L_{\rm LSS}$ is the diameter of the LSS and the last class corresponds to the covering space.
To classify the realizations, we test the following machine learning algorithms: random forests, extreme gradient boosting classifier (\texttt{XGBoost}), one-dimensional (1D) convolutional neural networks (CNNs), two-dimensional (2D) CNNs, and complex 2D CNNs. 
In each case we obtain the results for the algorithm trained and tested on randomly Wigner-rotated and non-rotated realizations. 
Additionally, we present a small set of results for harmonic-space realizations with the length scale of $L \gtrsim L_{\rm LSS}$ in order to set preliminary expectations for the effectiveness of machine learning for classifying large universes with non-trivial topologies. 

To be clear, $E_1$  with a fundamental domain of small size (i.e., $L \lesssim L_{\rm LSS}$) is observationally excluded \cite{Cornish:2003db,ShapiroKey:2006hm, Vaudrevange:2012da, WMAP:2003elm, WMAP:2012nax, Planck:2013okc, Planck:2015gmu,COMPACT:2022nsu}. 
However, our objective in this paper is not to extend current observational limits to larger fundamental domains, but to begin developing the machine learning techniques that may ultimately be necessary to do so, and to identify the challenges that must be addressed.
To this end, we do not make explicit use of the circles-in-the-sky signature, which does not extend to domains larger than the LSS. We also note that an analogous problem has been explored using a Bayesian likelihood approach (see, e.g., \rcite{Kunz:2006}), and while we do not explore such methods here, we find it valuable to compare the results obtained in this work with the corresponding results from the likelihood-based approaches. We compare the two different types of methods in \cref{section:discussion}.

We demonstrate that the  machine learning methods which we have tested are all effective at classifying harmonic-space realizations with $>99\%$ accuracy for realizations that have coordinate axes aligned with the edges of the cubic fundamental domain of the torus.
We find that the results depend significantly on whether the orientation of the torus is already known.
For the realistic case when realizations have been randomly rotated, we find that topologies with larger fundamental domains, e.g., with $L = 0.5 \times L_{\rm LSS}$, are more challenging to distinguish from the covering space.
Finally, we show that our methods are also effective when we classify non-rotated $a_{\ell m}$  $E_{1}$ realizations with the size scale slightly larger than the diameter of the LSS, i.e., for $L = 1.01 \times L_{\rm LSS}$, as long as the dataset is sufficiently large. 
We identify this difficulty in dealing with coordinate rotations as a key challenge to be addressed in future work.

The layout of the paper is as follows. In \cref{section:the_dataset} we discuss the theoretical background, the process of generating the $E_1$ topology realizations, and the general features of the resulting datasets. \cref{section:machine_learning_algorithms} outlines the machine learning algorithms we employ. \cref{subsection:results_small_E1} summarizes the results obtained for realizations with the size scale $L$ significantly smaller than that of the diameter of the LSS. The results for realizations with the size scale $L \gtrsim L_{\rm LSS}$ are discussed in \cref{subsection:results_L_L_LSS}. Future work, implications for observational topology searches, and a comparison with likelihood-based approaches are discussed in \cref{section:discussion}. \cref{appendix:A} contains extra information on the training procedures as well as the settings used to train the algorithms. The methodology for generating the realizations with $L \gtrsim L_{\rm LSS}$ is discussed in \cref{appendix:large_L_results}.

\section{The dataset}
\label{section:the_dataset}

\subsection{Properties of the $E_{1}$ topology}

To generate simulated realizations of the CMB on a manifold, we require the eigenmodes of the Laplacian on that manifold.\footnote{In principle, both the scalar (spin-zero) and tensor (massless spin-two) eigenmodes, but in this paper we confine ourselves to scalar perturbations.}
The eigenmodes for the Euclidean topologies have been studied extensively (e.g., see \rcite{COMPACT:2023_paper2A}). 
The $E_{1}$ eigenmodes $\Upsilon_{\boldsymbol{k}}^{E_1}$ are the subset of the $E_{18}$ eigenmodes that respect the $E_{1}$ symmetries,
\begin{equation}
    \label{eqn:E1eigenmodeinvariance}
\Upsilon_{\boldsymbol{k}}^{E_1}\left(g_{A_j}^{E_1} \boldsymbol{x}\right)=\Upsilon_{\boldsymbol{k}}^{E_1}(\boldsymbol{x}),\quad j=1,2,3\,.
\end{equation}

\noindent Here $g_{A_j}^{E_1}$ is a generator for the $E_{1}$ topology (described in detail in Sections 2 and 3.1 of \rcite{COMPACT:2023_paper2A})
i.e., a translation.
Since in this paper we will confine our attention to cubic $E_1$ manifolds,
these three translations are in orthogonal directions, which we can take to be aligned along the three coordinate  axes, and are of equal length, $L$.  
$L$ is therefore also the side length of the cubic periodic box that is the most convenient fundamental domain for cubic $E_1$. The symmetry condition \eqref{eqn:E1eigenmodeinvariance} causes the discretization of the allowed wave vectors.
In a cubic periodic box of side length $L$, this corresponds to
\begin{equation}  
    \boldsymbol{k}_{\boldsymbol{n}} =
    \frac{2 \pi}{L} (n_1, n_2, n_3),
\end{equation}
where the wave vectors $\boldsymbol{k}_{\boldsymbol{n}}$ are labeled by a triplet of integers $\boldsymbol{n}=\left(n_1, n_2, n_3\right)$.
The $E_{1}$ eigenmodes are therefore given by
\begin{equation}
\Upsilon_{\boldsymbol{k}_{\boldsymbol{n}}}^{E_1}(\boldsymbol{x})=e^{i \boldsymbol{k}_{\boldsymbol{n}} \cdot\left(\boldsymbol{x}-\boldsymbol{x}_0\right)}, \quad \text { for } \boldsymbol{n} \in \mathcal{N}^{E_1},
\end{equation}
\noindent where $\mathcal{N}^{E_1} \equiv\left\{\left(n_1, n_2, n_3\right) \mid n_i \in \mathbb{Z}\right\} \setminus (0,0,0)$. 

Given the discretization of the allowed wave vectors due to the $E_{1}$ symmetries, the spherical-harmonic coefficients can be obtained by
\begin{equation}
a_{\ell m}=\frac{4 \pi}{L^{3}} \sum_{\boldsymbol{n} \in \mathcal{N}^{E_{1}}} \delta_{\boldsymbol{k}_{\boldsymbol{n}}}^{\mathcal{R}} \xi_{k_{\boldsymbol{n}} \ell m}^{E_{1} ; \hat{\boldsymbol{k}}_{\boldsymbol{n}}} \Delta_{\ell}\left(k_{\boldsymbol{n}}\right),
\label{eq:a_lm_realizations}
\end{equation}

\noindent where $\xi_{k_{\boldsymbol{n}} \ell m}^{E_{1}; \hat{\boldsymbol{k}}_{\boldsymbol{n}}} \equiv e^{-i \boldsymbol{k}_{\boldsymbol{n}} \cdot \boldsymbol{x}_0} i^{\ell} Y_{\ell m}^*\left(\hat{\boldsymbol{k}}_{\boldsymbol{n}}\right)$ and $\delta_{\boldsymbol{k}_{\boldsymbol{n}}}^{\mathcal{R}}$ (primordial density fluctuations) are Gaussian, random, statistically independent variables 
of zero mean with  variances determined by the primordial power spectrum $\mathcal{P^{R}}$,
\begin{equation}
    \left\langle \delta_{\boldsymbol{k}_{\boldsymbol{n}}}^{\mathcal{R}} \delta_{\boldsymbol{k'}_{\boldsymbol{n}}}^{\mathcal{R}*} \right\rangle=\mathcal{P^{R}}(\boldsymbol{k_n})\delta^{(\mathrm{K})}_{\boldsymbol{k_n}\boldsymbol{k'_n}}\,.
\end{equation}
Note that $\Delta_{\ell}(k_{\boldsymbol{n}})$ in \cref{eq:a_lm_realizations} is the corresponding transfer function. The harmonic-space covariance matrix in the $E_{1}$ topology then has the form 

\begin{equation}
C_{\ell m \ell^{\prime} m^{\prime}}=\frac{(4 \pi)^2}{L^{3}} \sum_{\boldsymbol{n} \in \mathcal{N}^{E_{1}}} \Delta_{\ell}\left(k_{\boldsymbol{n}}\right) \Delta_{\ell^{\prime}}^{*}\left(k_{\boldsymbol{n}}\right) \frac{2 \pi^2 \mathcal{P}^{\mathcal{R}}\left(k_{\boldsymbol{n}}\right)}{k_{\boldsymbol{n}}^3} \xi_{k_{\boldsymbol{n}} \ell m}^{E_{1} ; \hat{\boldsymbol{k}}_{\boldsymbol{n}}} \xi_{k_{\boldsymbol{n}} \ell^{\prime} m^{\prime}}^{E_{1} ; \hat{\boldsymbol{k}}_{\boldsymbol{n}} *}.
\label{eq:analytic_cov_matrix}
\end{equation}

\noindent  In the rest of this work we will refer to the covariance matrix generated from individual $a_{\ell m}$ realizations as $\mathcal{C}_{\ell m \ell^{\prime} m^{\prime}} = a_{\ell m} a_{\ell^{\prime} m^{\prime} }^{*}$ as opposed to the analytic covariance matrix given in \cref{eq:analytic_cov_matrix}.

\subsection{Generating $a_{\ell m}$ realizations}

The $a_{\ell m}$ realizations can be generated by numerically evaluating \eqref{eq:a_lm_realizations} (following the approach previously applied in \rcite{COMPACT:2023_paper2A}).
In summary, we sum over all the Fourier modes that contribute to each spherical harmonic coefficient by generating the primordial density fluctuations, i.e., the $\delta_{\boldsymbol{k}_{\boldsymbol{n}}}^{\mathcal{R}}$, for each wave vector $\boldsymbol{k}_{\boldsymbol{n}}$. 
The transfer function along with the primordial power spectrum are obtained by using \texttt{CAMB} \cite{Lewis:1999bs,2011ascl.soft02026L} with {\it Planck} 2018 best-fit $\Lambda\textrm{CDM}$ cosmological parameters \cite{Planck:2018vyg}. 
Theoretically, we would need to perform the sums in the equations over an infinite set of wave numbers, however, in practice, we can obtain accurate results by choosing a cutoff value $k_{\max}$, at which we stop the summation. We choose $k_{\max}$ such that wave vectors with $k \leq k_{\max}(\ell)$ would contribute to at least 99\% of the $\Lambda$CDM angular power spectrum, $C_{\ell}^{\Lambda \mathrm{CDM}}$, as generated by \texttt{CAMB}.

We generate the $a_{\ell m}$ realizations for four classes of the $E_1$ topology.
We focus on small cubic $E_1$ manifolds and investigate the accuracy of different machine learning techniques in classifying the manifolds and distinguishing between them and the covering space. The chosen four classes are $L \in \{0.05, 0.1, 0.5, \infty \} \times L_{\rm LSS}$, where the last one corresponds to the $L \rightarrow \infty$ limit, i.e., the covering space. 
We expect that classifying realizations with $L > L_{\rm LSS}$ will be significantly more difficult. 
This is based on the fact that an observer in a universe of such size would not observe matched circle pairs in the CMB, so there are no (nearly) perfectly matched pixels (i.e., on a pixelated temperature map).

Similarly, the KL divergence and the signal-to-noise ratio statistic (introduced in Section 5.3 of \rcite{COMPACT:2023_paper2A}) for manifolds with $L > L_{\rm LSS}$ quickly approach unity \cite{COMPACT:2023_paper2A}. 
This indicates that classifying the realizations with $L > L_{\rm LSS}$ will likely require machine learning methods that are different from those outlined in this work. We plan to explore this in future work. Nevertheless, examining $L\leq1$ gives us an opportunity to explore the specific challenges we are likely to face in applying machine learning classification methods to cosmic topology.

In each case studied in the present work, the $a_{\ell m}$ realizations are generated with the multipoles in the range  $\ell \in [2,100]$.
However, for different machine learning algorithms we use different subsets of this dataset, depending on the memory constraints (for full details, see \cref{appendix:A}). 
In order to generate the covering space realizations, we use the \texttt{synalm} function available as part of the \texttt{healpy}\footnote{Available at \href{http://healpix.sourceforge.net}{http://healpix.sourceforge.net.}} package \cite{healpy1, healpy2}. 
We use the \textit{Planck} 2018 best-fit cosmological parameters (the same parameters as those used for computing the transfer function of the $E_1$ realizations). 
  
An important nuance when it comes to training machine learning models is  data ordering. 
By this we refer to the order of the $a_{\ell m}$ entries as stored in the data array. 
The order matters as it can determine whether certain features of the data appear locally or non-locally in the data vector. 
This is especially important for convolutional neural networks, where certain non-local features might not be captured depending on the size of the filters and the data arrays. To investigate such effects we consider two types of data orderings, $(m, \ell)$ and $(\ell, m)$. 
The $(m, \ell)$ ordering refers to the default \texttt{HEALPix} \cite{Gorski:2004by,Zonca2019} ordering, i.e., $a_{\ell m} = \{ a_{00},   a_{10}, a_{20}, \ldots, a_{ \ell_{max} 0}, a_{11}, a_{21}, \ldots, a_{\ell_{\mathrm{max}} \ell_{\mathrm{max}}}  \}$, while the $(\ell, m)$ ordering is $a_{\ell m}  = \{ a_{00}, a_{10}, a_{11}, a_{20}, a_{21}, \ldots, a_{\ell_{\mathrm{max}} \ell_{\mathrm{max}}} \}$. We denote a specific entry in a particular data vector by $\mathcal{I}_{\ell, m}$ for $(\ell, m)$ ordered dataset and by $\mathcal{I}_{m, \ell}$ for $(m, \ell)$ ordered dataset, where 

\begin{equation}
\mathcal{I}_{\ell, m}=\ell(\ell+1)/2+m, \hspace{10mm}\mathcal{I}_{m, \ell}=m(2\ell_{max}+1-m)/2+\ell.
\end{equation}

In order to train a 2D CNN, we require data formatted in 2D.
We generate the needed dataset by evaluating $\mathcal{C}_{\ell m \ell^{\prime} m^{\prime}}$ for each set of $a_{\ell m}$. 
The four-dimensional matrix is then flattened, such that for each 2D  $(\ell, \ell')$ ``block'' the corresponding $(m, m^{\prime})$ values are shown inside the block, resulting in the characteristic \textit{checkerboard} pattern seen (partially) in \cref{fig:cm_samples_norm} and (more clearly) in \cref{fig:cm_analytic_plots}. 
We generate the 2D data in the same two ordering schemes as are used for the $a_{\ell m}$ realizations. 
The $\mathcal{C}_{\ell m \ell^{\prime} m^{\prime}}$ data is generated for $\ell \in [2,20]$.
For larger $\ell_{\rm max}$, the dataset quickly becomes unmanageable. 
We expect this data format to be especially useful, as previous work indicates that for the larger ($L>L_{\rm LSS}$) manifolds that we eventually want to be able to classify the key features of non-trivial topologies for scalar temperature fluctuations are stored in the correlations between relatively low multipoles. 

An important choice when generating  the CMB realizations is the orientation of the coordinate system. 
We will find that this choice strongly affects the results of the machine learning classification.
To take this into consideration, we generate two sets of realizations, one where the $a_{\ell m}$ are unrotated compared to the natural frame of the cubic $E_1$ domain, and 
one where each realization is randomly rotated (i.e., by applying the Wigner D-matrices $D_{m \bar{m}}^{\ell}\left(\theta_0, \phi_0, \psi_{0}\right)$). 
In the unrotated case, the default orientation is such that the closest clone images are in the $x$, $y$, and $z$ directions, i.e., the observer's coordinate system is aligned along the three coordinate axes (see \rcite{wigner2013group, COMPACT:2023_paper2A} for a wider discussion of this point).

In order to prepare our training, validation, and test datasets, we numerically generate 40,000 $a_{\ell m}$ realizations, i.e., the same 10,000 realizations per class ordered in ($\ell$, $m$) and ($m$, $\ell$) ordering schemes (see \cref{section:machine_learning_algorithms} for further details). 
We also generate an augmented dataset in order to investigate the effects on the classification accuracy of an increased total dataset size and of rotations.
This is done by taking each of the aforementioned 40,000 realizations and randomly rotating it, choosing 
the three Euler angles so that the direction of the $z$ axis is drawn from a uniform distribution on the sky, and the orientation of the $x$ and $y$ axes with respect to that $z$ axis is drawn from a  uniform distribution from $0$ to $2\pi$. 
This is repeated 10 times for each of the initial 40,000 realizations, resulting in a total of 400,000 realizations.
Different subsets of this augmented dataset are then used to train the different algorithms, as outlined in \cref{appendix:A}.  

In addition to the datasets described above, we generate an extra set of $a_{\ell m}$ realizations with $L \gtrsim L_{\rm LSS}$ by employing an alternative method of Cholesky decomposition (see \rcite{watkins2004, Mukherjee2014, COMPACT:2023_paper2A}).
The primary motivation for using an alternative method for generating the data here is the fact that numerically evaluating \cref{eq:a_lm_realizations} for topology scales of $L \gtrsim L_{\rm LSS}$ is not viable given the computational resources available to us. 
An alternative method for generating harmonic-space realizations is that of computing the temperature auto-correlation matrix, for given values of $L$ and $\ell_{\rm max}$, and then using Cholesky decomposition to factor the correlation matrix into a lower triangular matrix along with its conjugate. 
This method allows generating harmonic-space realizations corresponding to the topology size scale $L \gtrsim L_{\rm LSS}$ significantly faster, with the caveat that the maximum multipole value is lower, e.g., $\ell_{\rm max} \approx 30$. 
This is not an issue, since we already know \cite{COMPACT:2023_paper2A, Fabre2015} that for $L>L_{\rm LSS}$ most of the information in the CMB for distinguishing compact topologies from the covering space is at  $\ell\lesssim 30$.

Specifically, we generate a number of datasets for two topology classes, cubic $E_{1}$ topology with  $L = 1.01 \times L_{\rm LSS}$ and the covering space. 
This particular size scale is chosen as a test case for a manifold that is sufficiently large for there to be no matched circle pairs, but not too large (as we know that for $L \gtrsim 1.1\times L_{\rm LSS}$ the KL divergence and the signal-to-noise ratio would be significantly small \cite{COMPACT:2023_paper2A}, and hence, an alternative classification approach would likely be required). The realizations are generated with $\ell_{\mathrm{max}} = 30$. 
In order to investigate how the classification accuracy depends on the dataset size, a total of 6 datasets are generated, ranging in their total number of realizations between 200 and 200,000. 
As before, we generate two versions of each dataset -- an unrotated and a randomly rotated one. In each case we chose to work with the realizations in the $(\ell, m)$ ordering. Further technical details, as well as a comparison between the realizations obtained by the two outlined methods, are described in \cref{appendix:large_L_results}.

\subsection{Features of the $a_{\ell m}$ realization data}

\begin{figure}
  \centering
  \includegraphics[width=0.95\textwidth]{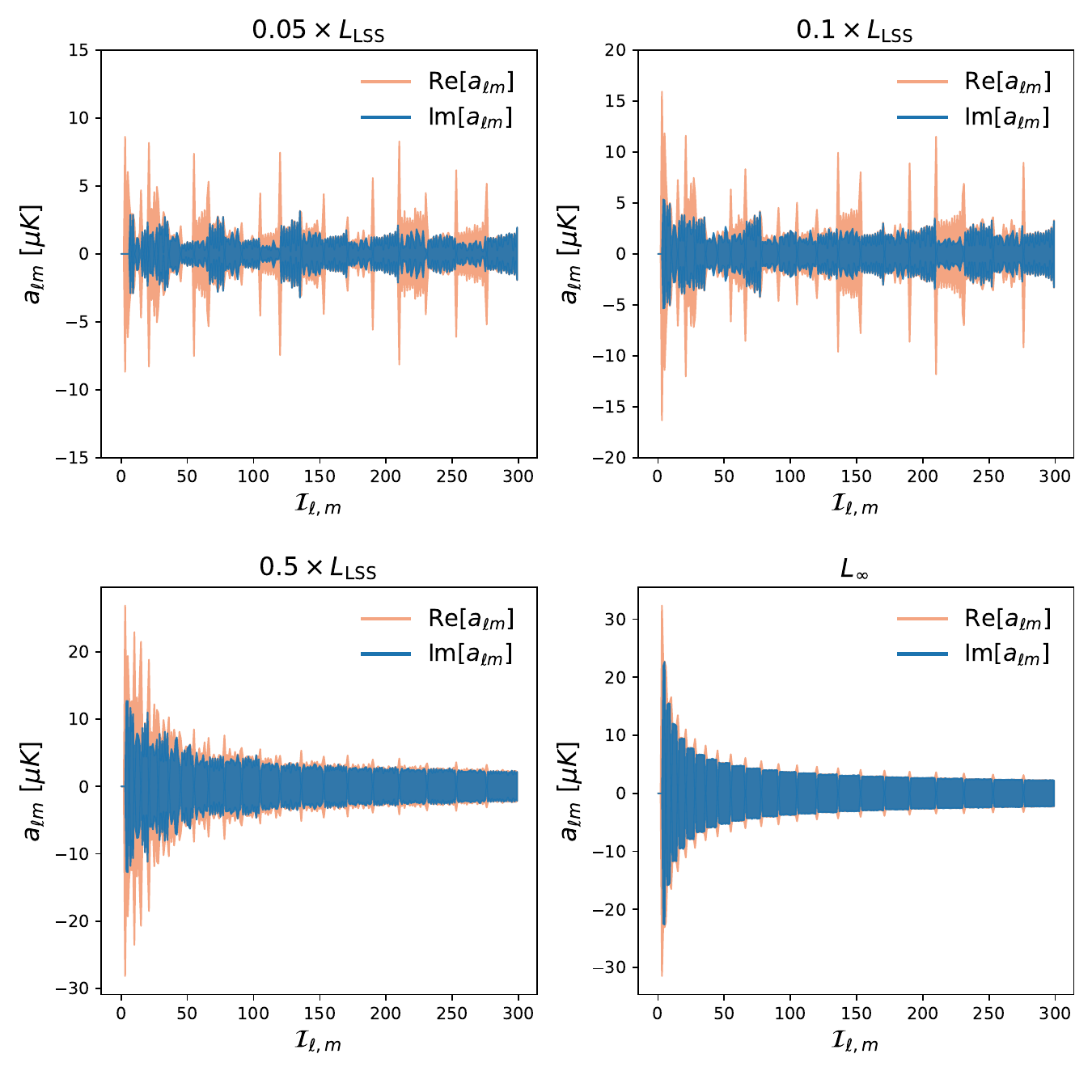}
  \caption{Samples of each dataset class, i.e., $L \in \{ 0.05, 0.1, 0.5, \infty  \} \times L_{\rm LSS}$, in $(\ell, m)$ ordering for the cubic $E_1$ topology. For each class, the mean and the standard deviation over the full dataset (10,000 realizations) are plotted and we show the real and imaginary parts of the $a_{\ell m}$ for the first 300 values in order to illustrate the local features of the dataset. The $L_{\infty}$ class corresponds to the covering space (or trivial topology).}
  \label{fig:a_lm_samples_l_ordering}
\end{figure}

\begin{figure}
  \centering
  \includegraphics[width=0.95\textwidth]{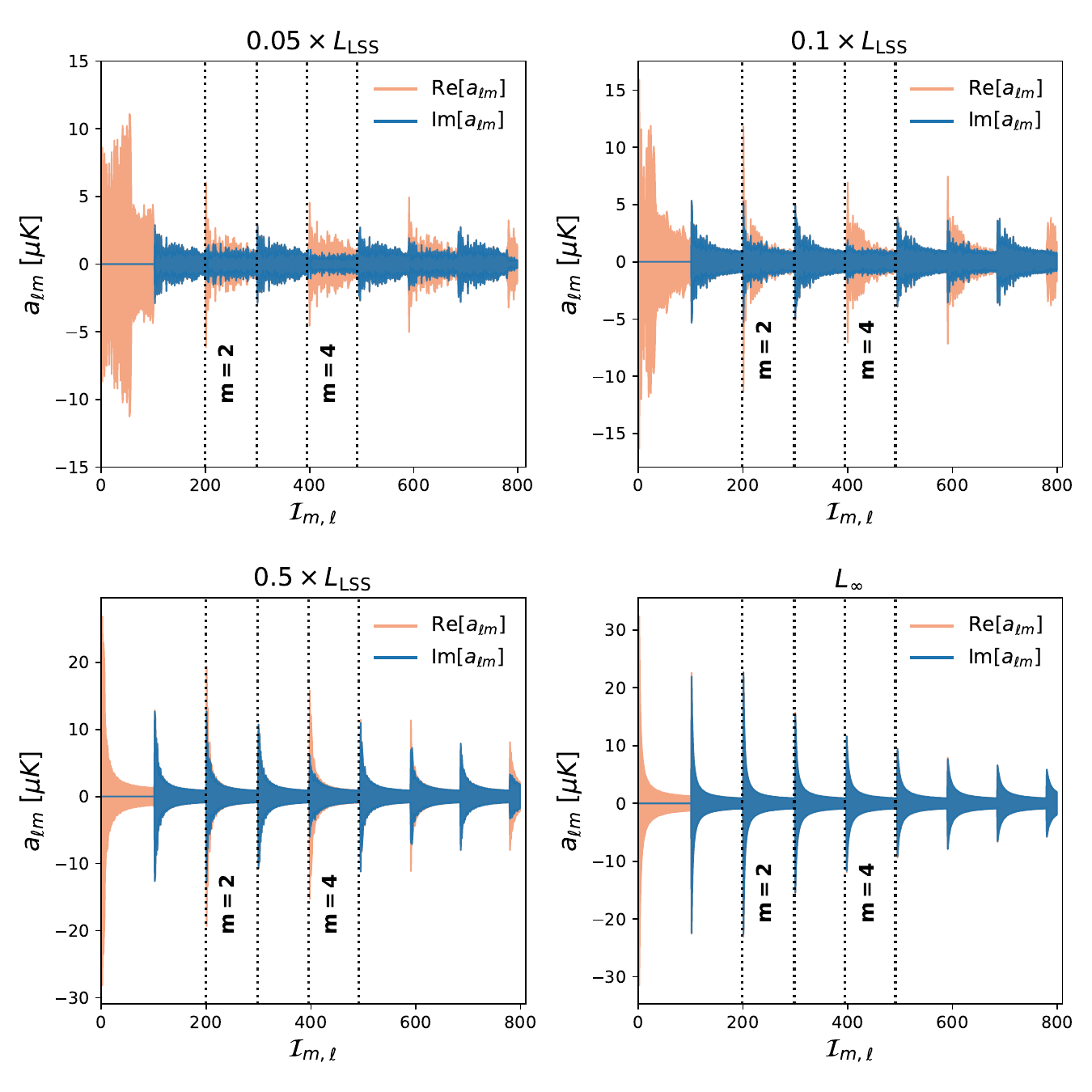}
  \caption{As in \cref{fig:a_lm_samples_l_ordering}, but for $(m, \ell)$ ordering. For each sample we show the first several $m$ values. In particular, $m = 2$ and $m = 4$ are marked by dashed vertical lines to demonstrate the suppression of the imaginary values with respect to the corresponding real values.}
  \label{fig:a_lm_samples_m_ordering}
\end{figure}

Here we lay out some of the key features of the generated datasets. \cref{fig:a_lm_samples_l_ordering} shows the mean and the standard deviation for each class in the dataset.
For each class real and imaginary parts are shown in ($\ell$, $m$) ordering.
Similarly, \cref{fig:a_lm_samples_m_ordering} shows random data samples in $(m, \ell)$ ordering.
The first two even $m$ values are marked in the figure for each class.
As expected, we find that larger $E_{1}$ realizations, e.g., realizations with $L = 0.5 \times L_{\rm LSS}$, look nearly identical to the covering space realizations.
Meanwhile, smaller realizations, such as $L = 0.05 \times L_{\rm LSS}$, show clear differences when compared to the trivial topology class.
One such feature is the visible suppression of the variance of the imaginary parts for even $m$ values.
The same effect is seen even more clearly in \cref{fig:a_lm_samples_l_ordering}.
This feature is more noticeable in small $E_{1}$ ($L = \{0.05-0.1\}\times L_{\rm LSS}$) realizations, less visible for the larger $E_{1}$ realizations ($L=0.5 \times L_{\rm LSS}$), and not present at all in the covering space realizations.
Note that this is a feature that may be specific to cubic $E_{1}$ and we do not expect it to necessarily be present in non-cubic, randomly rotated realizations or in other topologies.

\begin{figure}
    \centering

    \begin{subfigure}{0.45\textwidth}
        \centering
        \includegraphics[width=\linewidth]{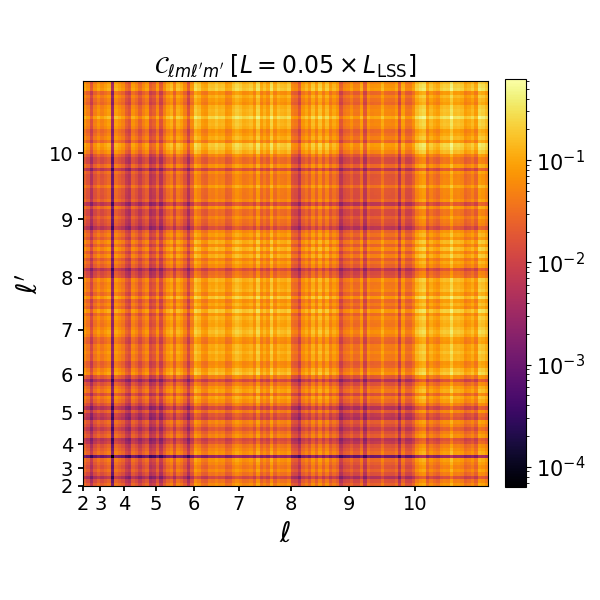}
        \label{fig3:sub1}
    \end{subfigure}
    \begin{subfigure}{0.45\textwidth}
        \centering
        \includegraphics[width=\linewidth]{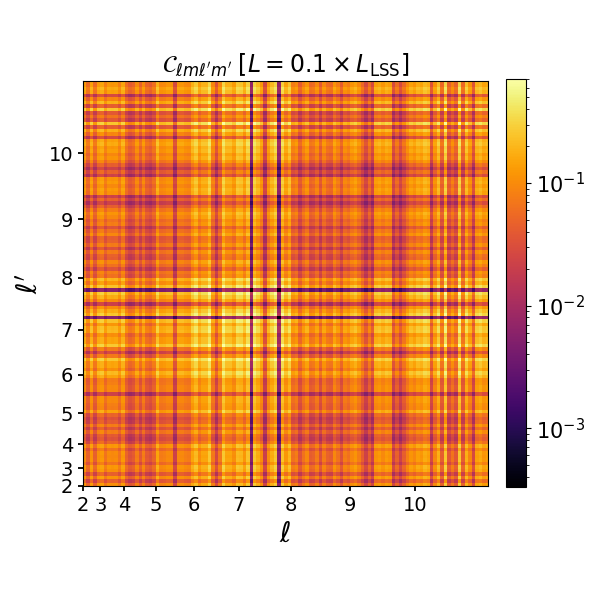}
        \label{fig3:sub2}
    \end{subfigure}

   \vspace{-1.1cm}

    \begin{subfigure}{0.45\textwidth}
        \centering
        \includegraphics[width=\linewidth]{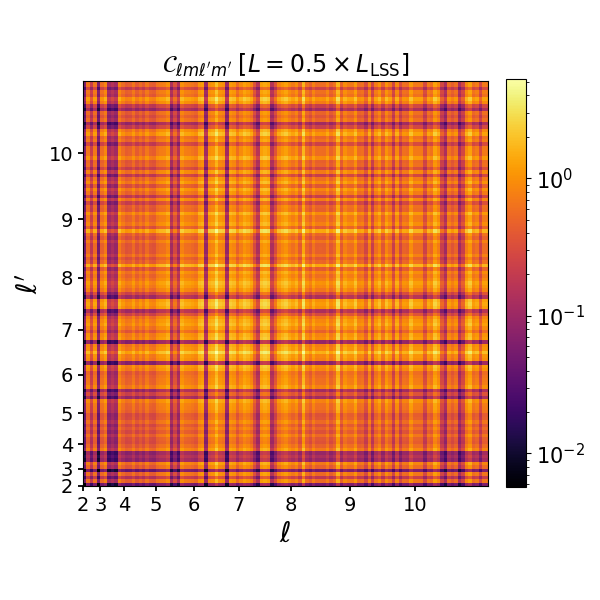}
        \label{fig3:sub3}
    \end{subfigure}
    \begin{subfigure}{0.45\textwidth}
        \centering
        \includegraphics[width=\linewidth]{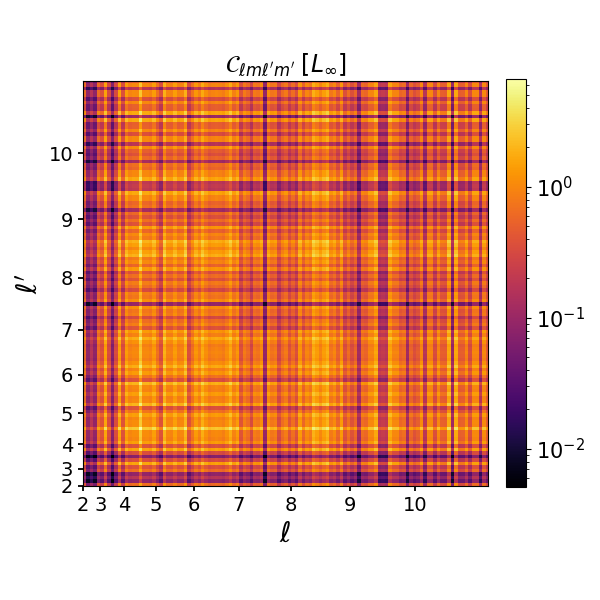}
        \label{fig3:sub4}
    \end{subfigure}
    \vspace{-1.0cm}
    \caption{A selection of random samples of the absolute value of $\mathcal{C}_{\ell m \ell^{\prime} m^{\prime}}$ (represented as a 2D array) for each class in our dataset. For each $(\ell, \ell')$ block the matrix elements show the corresponding $(m, m')$ in increasing order, i.e., $-\ell \leq m \leq \ell$ (see Fig. \ref{fig:cm_analytic_plots} for a clearer view of the block structure). Each plot is rescaled by $\left(C_{\ell}^{\Lambda \mathrm{CDM}}C_{\ell^{\prime}}^{\Lambda \mathrm{CDM}}\right)^{-1/2}$,
    with $C_{\ell}^{\Lambda \mathrm{CDM}}$ the $\Lambda \mathrm{CDM}$ angular power spectrum. }
    \label{fig:cm_samples_norm}
\end{figure}

\begin{figure}
    \centering

    \begin{subfigure}{0.45\textwidth}
        \centering
        \includegraphics[width=\linewidth]{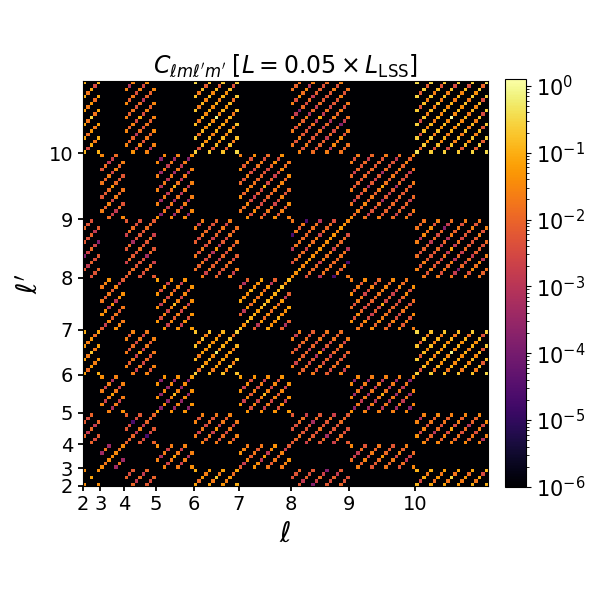}
        \label{fig4:sub1}
    \end{subfigure}
    \begin{subfigure}{0.45\textwidth}
        \centering
        \includegraphics[width=\linewidth]{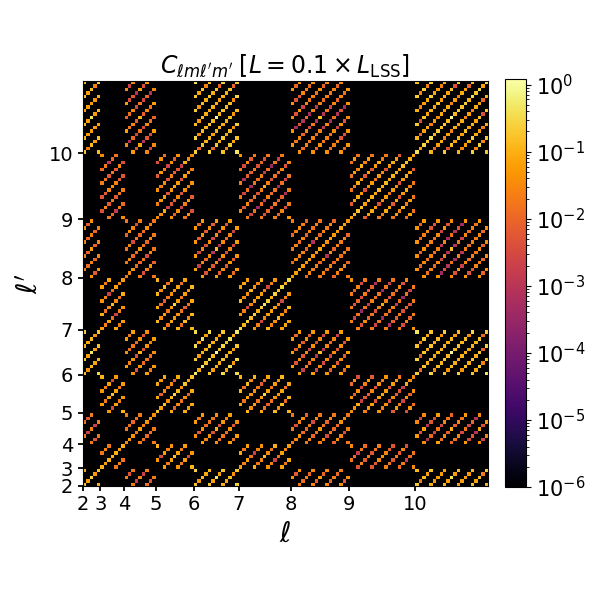}
        \label{fig4:sub2}
    \end{subfigure}

    \vspace{-1.1cm}

    \begin{subfigure}{0.45\textwidth}
        \centering
        \includegraphics[width=\linewidth]{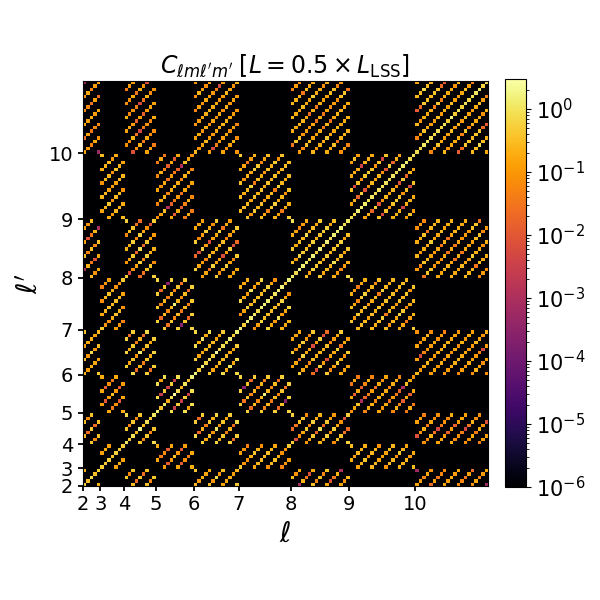}
        \label{fig4:sub3}
    \end{subfigure}
    \begin{subfigure}{0.45\textwidth}
        \centering
        \includegraphics[width=\linewidth]{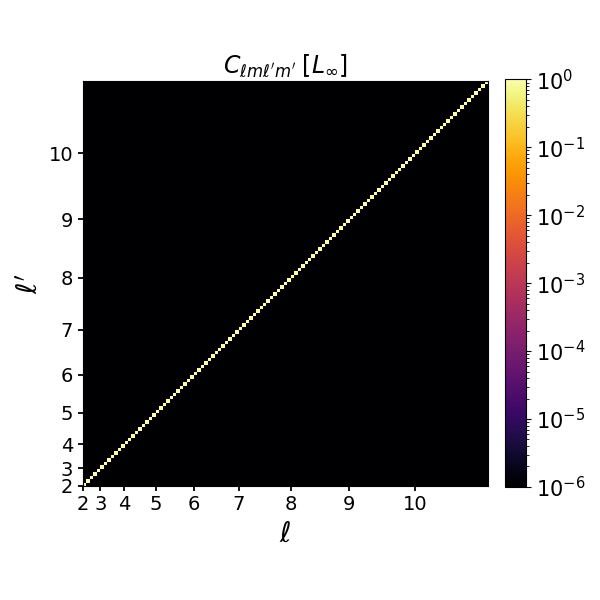}
        \label{fig4:sub4}
    \end{subfigure}
    \vspace{-1.0cm}

    \caption{The analytic covariance matrix \eqref{eq:analytic_cov_matrix} for each class in our dataset. The absolute value is shown in each case, and each covariance matrix is rescaled by 
    $\left(C_{\ell}^{\Lambda \mathrm{CDM}}C_{\ell^{\prime}}^{\Lambda \mathrm{CDM}}\right)^{-1/2}$.}
    \label{fig:cm_analytic_plots}
\end{figure}

\cref{fig:cm_samples_norm} shows a selection of randomly chosen $\mathcal{C}_{\ell m \ell^{\prime} m^{\prime}}$ realizations, while \cref{fig:cm_analytic_plots} shows the analytic covariance matrices, as defined in \cref{eq:analytic_cov_matrix}, for different sizes of the $E_{1}$ topology. 
The figures all show the absolute values of correlation normalized by the $\Lambda \text{CDM}$ power spectrum, i.e., by $\left(C_{\ell}^{\Lambda \mathrm{CDM}} C_{\ell^{\prime}}^{\Lambda \mathrm{CDM}}\right)^{-1/2}$. 
The key standout feature is the \textit{checkerboard} pattern observed in \cref{fig:cm_analytic_plots}, which indicates off-diagonal correlations between different  multipoles. 
The relatively orderly and local 
distribution of features (i.e., nearby in the $C_{\ell m \ell' m'}$ matrix given the $(\ell,m)$ ordering) hints at this data format being advantageous over some other ways of representing the data.
For example, one could consider training an algorithm directly on the temperature fluctuation \texttt{HEALPix} maps for each topology class (for instance by using \texttt{DeepSphere} \cite{Defferand2020}). While it is true that for the small topology scales that we are considering here (i.e., $L < L_{\rm LSS}$) the correlations might be clearly represented in the real (or pixel) space, extracting this information is non-trivial. It generally requires specific algorithms (e.g., for finding the matched-circle pairs) and a sufficiently large value of $\ell_{\rm max}$ (generally at least $\ell_{\rm max} \approx 250-300$).
Extracting this information with machine learning is also complicated by the fact that the correlated pixels generally appear non-locally in the map and are difficult to capture by the convolutional filters. 
This is especially true in the regime of $L > L_{\rm LSS}$ (which is our ultimate target given the existing negative circle-search results), where we expect correlations to be the clearest in (three-dimensional) Fourier space, and hence, in harmonic space. 
For these two reasons, we choose to work specifically with harmonic space realizations with the maximum multipole value in the range of $\ell_{\rm max} = 50 - 100$ (and $\mathcal{C}_{\ell m \ell^{\prime}m^{\prime}}$ realizations with $\ell_{\rm max} = 20$).

\section{Machine learning algorithms}
\label{section:machine_learning_algorithms}

To classify the different realizations of the covering space and the $E_{1}$ topology we use a number of different machine learning algorithms. 
Here we summarize the key features of those algorithms along with the details of the training procedure for each case. Further technical details on the training procedure and the data preparation are provided in \cref{appendix:A}.

\subsection{Random forests and extreme gradient boosting classifier trained on $a_{\ell m}$ data}
\label{section:random_forests}

\emph{Random forests} refers to an ensemble algorithm that is widely used for classification and regression problems.
Specifically, it is an ensemble of individual decision trees, the predictions of which are combined using ensemble voting. 
The main features of the algorithm along with common uses are described in detail in \rcite{Biau2010, Genuer2015, Biau2015}.
During the training procedure, a random subset of the training data vectors is sampled (with replacement). 
For a given subset of the data, a decision tree is generated by determining the optimal way to split the data values, such that the classification accuracy is maximized. 
The optimal splits are evaluated based on a chosen statistic, generally Gini impurity, entropy, or mean squared error for regression tasks. Each tree is constructed until the specified stopping condition is met (e.g., maximum depth of the tree or a minimum number of samples in a leaf node). 
Once the specified number of trees are generated, their predictions are combined in an ensemble vote. 
This is generally a majority vote or an average over all predictions. 

Random forests correct some of the common issues observed when using decision trees, namely their tendency to overfit when the number of features is high. 
This is an advantage for our dataset, where the number of features (i.e., $a_{\ell m}$ entries) is comparable to the size of the dataset (i.e, the number of realizations in the training dataset). 
In addition, it is generally found that multiple uncorrelated models (i.e., random forests) perform significantly better for classification tasks than individual models (i.e., individual decision trees) do. Random forests are also known to be more interpretable than some of the well-known \textit{black box} algorithms, e.g., neural networks. 
As a concrete example, one can calculate the feature importance for a given feature in our dataset. 
This refers to a Gini impurity measure of how often a randomly chosen element would be incorrectly classified. 
For a given node in a random forest classifier, the Gini impurity refers to the uncertainty of that particular node with respect to the target variable. 
It can be expressed as $
G =1-\sum_{i=1}^C p_i^2$, with $C$ the number of classes and $p_{i}$ the probability of class $i$ in that particular node. 
The feature importance is then calculated as the sum of the Gini impurity decrease over all nodes in all decision trees where the feature at hand was used to make a classification decision. 
For our specific dataset of $a_{\ell m}$ realizations, the feature importance score then tells us which particular values of $\ell$ and $m$ are most significant when making the classification decision.
In other words, the feature importance statistic  indicates which angular scales on the CMB sky hold the key information that differentiates non-trivial topology realizations from one another and from the corresponding covering space realizations. 

In this work, we train a random forest classifier available from the \texttt{scikit-learn} library \cite{scikit-learn}. 
The training is performed on both the randomly rotated (100,000 data samples) and the unrotated samples (a dataset of 40,000 samples). 
For both datasets, we consider $a_{\ell m}$ values for $\ell \in [2, 100]$. 
The data is split into 80\% training and 20\% test datasets. 
The training is performed with 3500 estimators (number of decision tress). 
For the split criterion, we use Gini impurity. 
The data samples are prepared by splitting the complex $a_{\ell m}$ into their real and imaginary parts (in $(\ell, m)$ ordering), 
which are appended into a single data vector. 
In addition to the real and imaginary parts of each realization, we consider many other features that can be constructed from the available data, e.g., $C_{\ell}$ values, average $a_{\ell m}$ values for a given $\ell$, and the eigenvalues of the $\mathcal{C}_{\ell m \ell^{\prime} m^{\prime}}$ matrix corresponding to a given realization. 
We find that the eigenvalues of $\mathcal{C}_{\ell m \ell^{\prime} m^{\prime}}$ are particularly informative. 
This is the case, as, given the structure of the $a_{\ell m} a^*_{\ell'm'}$ matrix for a given realization, all but one of the eigenvalues are zero.
The only non-zero eigenvalue corresponds to the square of the absolute value of the $a_{\ell m}$ data vector, i.e., $
\vert\vert a_{\ell m}\vert\vert^2 \equiv\sum_{\ell m}\vert a_{\ell m}\vert^2$, which is a rotation-invariant quantity. 
Hence, in order to improve the classification results, we append $\vert\vert a_{\ell m}\vert\vert^2$ to the training and test data arrays. 
Further training parameter settings and the motivation behind choosing their values are outlined in \cref{appendix:random_forests}. 

A class of algorithms that share many similarities with random forests is that of \emph{gradient boosting}.
Gradient boosting here specifically refers to combining multiple models trained sequentially such that every new model corrects the classification errors made by the existing ensemble. 
In practice, gradient boosting is generally implemented using multiple weak learners, such as decision trees. 
The initial tree is generated by choosing the most important features of the dataset based on the decrease of a chosen loss function. The feature with the highest loss decrease then becomes the root node. 
Additional nodes of the tree are then added iteratively based on their loss value until the specified maximum depth of the tree is reached. 
The outer layers of the tree (leaf nodes) then hold the prediction of the tree and can be used to calculate the residual values by comparing to the data labels in the test dataset. 
The subsequent trees during the training procedure are trained in an analogous way, the only difference being that they are trained on the residual data. 
This feature, in particular, allows the algorithm to correct the errors made during the previous training iteration. 

A particularly useful implementation of gradient boosting is offered by the Python package \texttt{XGBoost} \cite{XGBoost_paper, xgboost_documentation}. 
In recent years, \texttt{XGBoost} has been used to generate state-of-the-art results in classification and regression for many astrophysics applications, often producing similar or better results than those obtained with commonly used neural network algorithms. 
In addition, \texttt{XGBoost} has been used as a winning algorithm for multiple Kaggle competitions.
Being based on ensembles of decision trees, the predictions made by \texttt{XGBoost} are generally not prone to overfitting for a large number of features. 
Similarly, having an ensemble of decision trees allows us to calculate the relative feature importance in a fashion analogous to the random forest algorithm.  

We train an extreme gradient boosting classifier provided by the \texttt{XGBClassifier} class from the \texttt{XGBoost} Python package. 
The training settings for the classifier are generally analogous to the one used for random forests, with the main exception being that we use 2000 decision tree classifiers (rather than 3500 for the random forest classifier). 
The data is split into 80\% training and 20\% test datasets. The complex values are split into the real and imaginary parts, which are appended into a single data vector. 
The non-zero eigenvalue of the $\mathcal{C}_{\ell m \ell^{\prime} m^{\prime}}$ is also appended to each data vector. The full list of settings used, along with how the particular values are chosen, is described in \cref{appendix:random_forests}.

\subsection{1D convolutional neural networks trained on $a_{\ell m}$ data}

CNNs refer to the class of neural networks that employ convolutions in order to extract features from N-dimensional data.
The key building blocks of CNN algorithms are the convolutional layers, which apply convolution operations to the input data. Specifically, these layers include filters or kernels of a chosen size that slide across the input data and perform convolutions, which extract a set of feature maps.
Extra features can then be extracted in the subsequent layers, which are ultimately combined in the final set of (fully connected) layers of the network in order to generate a prediction. In addition to the mentioned convolutional layers, the networks often include extra elements, such as pooling layers that downsample the feature maps in order to reduce complexity and to extract the key information. 
Dropout layers are also often added to randomly switch off a certain fraction of neurons in order to avoid overfitting. 

1D CNNs refer to the sub-class of neural networks specifically applied to 1D datasets by performing 1D convolution operations. 
Due to the structure of 1D CNNs, it is an algorithm of choice for many signal processing tasks ranging from audio classification to astrophysical signal processing. 
Similarly, 1D CNNs have been widely applied for the analysis of sequential data, such as for speech recognition, time-series analysis, and natural-language processing.
In general, 1D CNNs share mostly the same structure as their 2D counterparts, the main difference being that the layers are strictly one-dimensional (\texttt{Conv1D}, \texttt{MaxPooling1D}, \textit{etc}).
Extra layers are sometimes added specifically for the analysis of sequential or time-series data, e.g., long-short-term-memory (LSTM) layers. 

We train the 1D CNN algorithm on the two mentioned datasets: the 40,000 unrotated $a_{\ell m}$ realizations and the 100,000 randomly rotated (and augmented via extra rotations) data samples.
For each realization, we extract the $a_{\ell m}$ data corresponding to $\ell \in [2,50]$ in $(\ell, m)$ ordering (see \cref{appendix:1D_CNN_architecture} for a discussion of why these particular values are chosen).
In order to deal with complex values of the $a_{\ell m}$ realizations, we  split the $a_{\ell m}$ into the real and imaginary parts, which are appended into a single array.
In addition, for the rotated dataset, the eigenvalues of the $\mathcal{C}_{\ell m \ell^{\prime} m^{\prime}}$ matrix corresponding to each $a_{\ell m}$ realization vector are also appended to the data array (see \cref{appendix:1D_CNN_architecture} for further technical details).
 
We build and implement the 1D CNN architecture using \texttt{TensorFlow} with \texttt{Keras} \cite{tensorflow2015-whitepaper, chollet2015keras}. 
Generally, we expect the key information related to non-trivial topologies to be stored as correlations between the $a_{\ell m}$ entries for different $\ell$'s and $m$'s.
In order to capture this information, we choose an architecture with multiple kernels of different sizes (see \cref{table:1D_CNN_architecture}).
In addition to the 1D convolution layers, we also add a down-sampling layer (\texttt{MaxPool1D}) followed by a dropout layer (with a dropout scale of 0.3) to reduce overfitting. 
The three final layers are fully connected dense layers, which combine the extracted features into a final classification prediction.
The model is compiled using the \texttt{Adam} optimizer along with the sparse categorical cross entropy loss function.
The model is trained for 40 epochs, which we find to be sufficient for the model to converge.
The weights corresponding to the model with the highest validation accuracy during the training procedure are saved as the best model. 
The test accuracy along with the confusion matrix are calculated for the full test dataset.

\subsection{2D convolutional neural networks trained on $\mathcal{C}_{\ell m \ell^{\prime} m^{\prime}}$ data}

Previous results in the literature indicate that one of the key signatures of non-trivial topology is the non-local correlation pattern between different $\ell$'s \cite{Fabre:2013wia, COMPACT:2022nsu, COMPACT:2022gbl, COMPACT:2023_paper2A}.
Generally, we expect this information to be easier to capture in two dimensions, i.e., by analyzing the auto-correlation data.
In ($\ell , m$) ordering, this results in 2D correlations resembling a checkerboard pattern seen in \cref{fig:cm_samples_norm}.
While we do expect that the correlations are more easily captured in two dimensions, there is a natural trade-off due to memory constraints -- we are limited to rather low values of $\ell_{\rm max}$, i.e., $\ell_{\rm max} \lesssim 20-30$.
This is the case as, even for $\ell_{\rm max} = 20$, the 2D $\mathcal{C}_{\ell m \ell^{\prime} m^{\prime}}$  array is significantly larger than the 1D vector of $a_{\ell m}$  with even $\ell_{\rm max} = 100$ ($42849$ data entries per realization versus 5151). 

We prepare the data by calculating  a 2D realization  $\{\mathcal{C}_{\ell m \ell^{\prime} m^{\prime}}\}$ for $\ell,\ell' \in [2, 20]$, for each 1D realization $\{a_{\ell m}\}$.
Following the approach described in \rcite{COMPACT:2023_paper2A}, each 2D realization is rescaled by $\left(C_{\ell}^{\Lambda \mathrm{CDM}}C_{\ell^{\prime}}^{\Lambda \mathrm{CDM}}\right)^{-1/2}$, with $C_{\ell}^{\Lambda \mathrm{CDM}}$  the $\Lambda \mathrm{CDM}$ covering-space angular power spectrum. 
Since the resulting correlation values are complex, we store the real and imaginary parts as different channels.
In addition to the imaginary and real values, we add an extra data channel corresponding to the absolute values of $\mathcal{C}_{\ell m \ell^{\prime} m^{\prime}}$, resulting in each data sample having the shape  $(207, 207, 3)$.
Due to memory constraints, we perform the training on the original dataset of 40,000 unrotated and the 40,000 randomly Wigner-rotated realizations (without data augmentation).

In an attempt to capture the correlations between the different multipoles, we choose to work with the \texttt{ResNet-50} architecture available from the \texttt{Keras} model library \cite{Resnet-50-2015}.
Originally designed by Microsoft Research, \texttt{ResNet-50} has demonstrated state-of-the-art results for image classification tasks.
Due to the use of convolutional layers with different filter sizes and pooling operations, \texttt{ResNet-50} based architectures are known to work well with data having features that are correlated at different scales.
Another advantage of the \texttt{ResNet-50} architecture is that the model available on \texttt{Keras} library allows working with pre-trained model weights obtained by training on the \texttt{ImageNet} dataset\footnote{Available at \href{https://image-net.org/index.php}{https://image-net.org/index.php}.}, which generally results in a faster training procedure. 
A more in-depth discussion of the \texttt{ResNet-50} architecture and the settings of the training procedure are available in \cref{appendix:2D_CNN_ResNet-50}.

\subsection{2D complex convolutional neural networks trained on $\mathcal{C}_{\ell m \ell^{\prime} m^{\prime}}$ data}

Both the coefficients $a_{\ell m}$ of the expansion \eqref{eq:Spherical} and the elements of the covariance matrix \eqref{eq:analytic_cov_matrix} are complex. 
In the previous sections, we split the complex inputs into real and imaginary parts. 
In this section, we  explore the use of complex-valued neural networks (CVNNs) \cite{Barrachina2021}, a set of neural networks capable of handling complex numbers.
It has been proven that CVNNs have a stronger generalization power than real-valued neural networks (RVNNs) \cite{Hirose2012}.
The findings in \rcite{barrachina2021complexvalued} indicate that CVNNs outperform RVNNs across various architectures and data structures. The authors also claim that the accuracy of CVNNs demonstrates a statistically higher mean and median, coupled with lower variance compared to RVNNs, and that when no regularization technique is applied, CVNNs exhibit reduced overfitting.

The CVNN library  \cite{Barrachina2021} generalizes the RVNNs using
\texttt{Tensorflow} as back-end to enable the use of complex inputs, weights, and activation functions.
The architecture used in the present work is based on the one described in Ref. \cite{Ribli:2019wtw}, where it was used to extract the cosmological parameters $\Omega_\mathrm{M}$ and $\sigma_8$ from noisy weak-lensing maps. 
The network consists of 15 2D convolutional complex layers grouped in sets of 2, 3, and 5 layers, with different kernel sizes, followed by a 2D average pooling layer in between sets (see \cref{table:CVNN_architecture}).
Lastly, two dense layers output the prediction for each input realization. The architecture is described in more detail in \cref{appendix:CVNN}.

The dataset used to train the CVNNs is the same as described in the previous section, i.e., $\mathcal{C}_{\ell m \ell^{\prime} m^{\prime}}$ data with $\ell \in [2, 20 ]$.
However, we use the complex array itself as input to the network. 
Due to memory constraints, we do not apply any data augmentation via extra rotation when training the CVNN (i.e., we use the rotated and unrotated data samples with 40,000 realizations in each case).
The data set is split into $80\%$ for training, $10\%$ for validation, and $10\%$ for test. 
The network is trained for 120 epochs using \emph{stochastic gradient descent} with a learning rate of 0.002 (see \cref{appendix:CVNN} for a further discussion of the chosen training parameters).

\section{Results}
\label{section:results}

The obtained results indicate that the classification accuracy depends on several factors, including the range of multipoles included in the dataset, the data ordering, and, most importantly, whether or not the realizations have been randomly rotated. 
Similarly, the results depend strongly on the topology scale.
In this section, we summarize the main findings for each of the algorithms considered for realizations with topology scale  $L < L_{\rm LSS}$. 
In addition, we present a set of results for realizations with $L \gtrsim L_{\rm LSS}$.

\subsection{$E_{1}$ realizations with $L < L_{\rm LSS}$}
\label{subsection:results_small_E1}

\begin{figure}
    \centering

    \begin{subfigure}{0.45\textwidth}
        \centering
        \includegraphics[width=\linewidth]{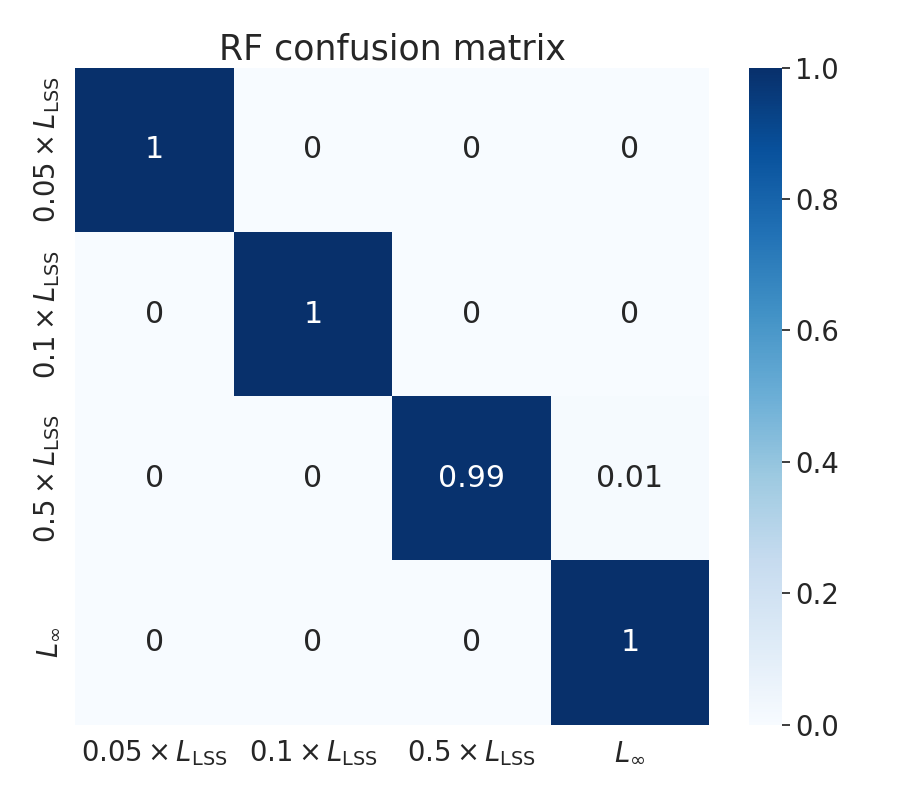}
        \label{fig_CM_RF_XGB:sub1}
    \end{subfigure}
    \begin{subfigure}{0.45\textwidth}
        \centering
        \includegraphics[width=\linewidth]{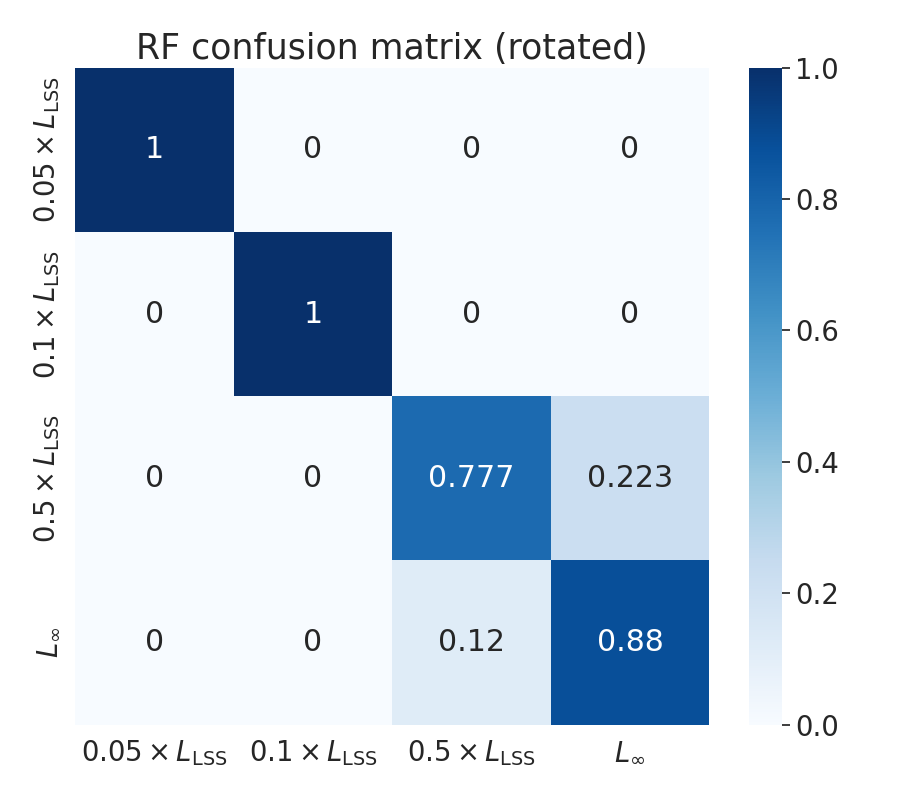}
        \label{fig_CM_RF_XGB:sub2}
    \end{subfigure}

    \medskip

    \begin{subfigure}{0.45\textwidth}
        \centering
        \includegraphics[width=\linewidth]{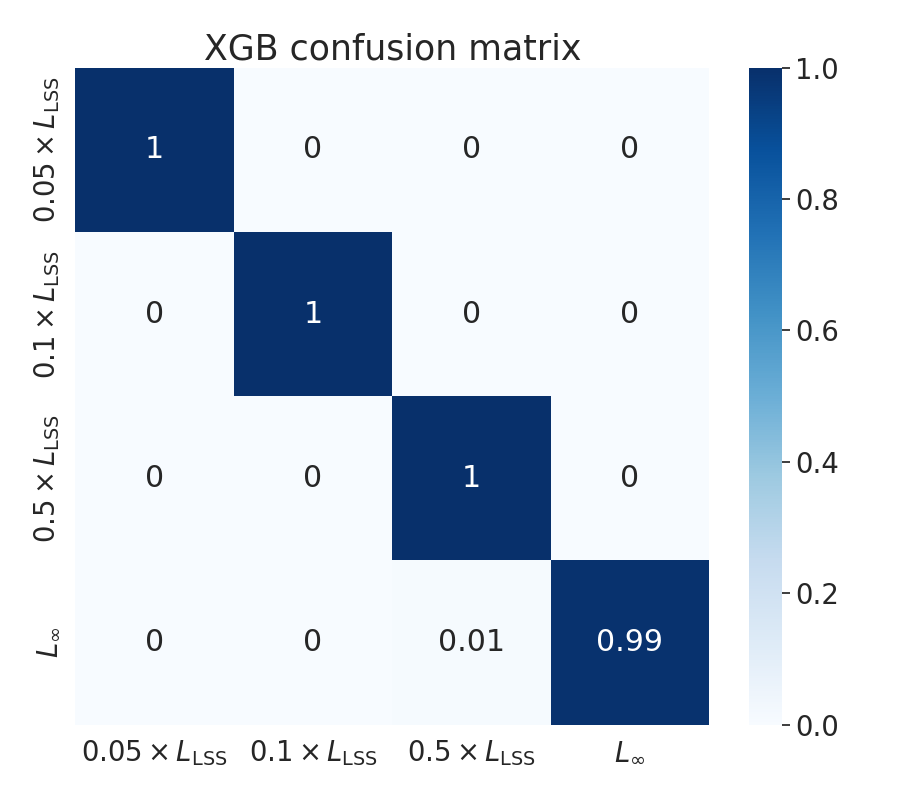}
        \label{fig_CM_RF_XGB:sub3}
    \end{subfigure}
    \begin{subfigure}{0.45\textwidth}
        \centering
        \includegraphics[width=\linewidth]{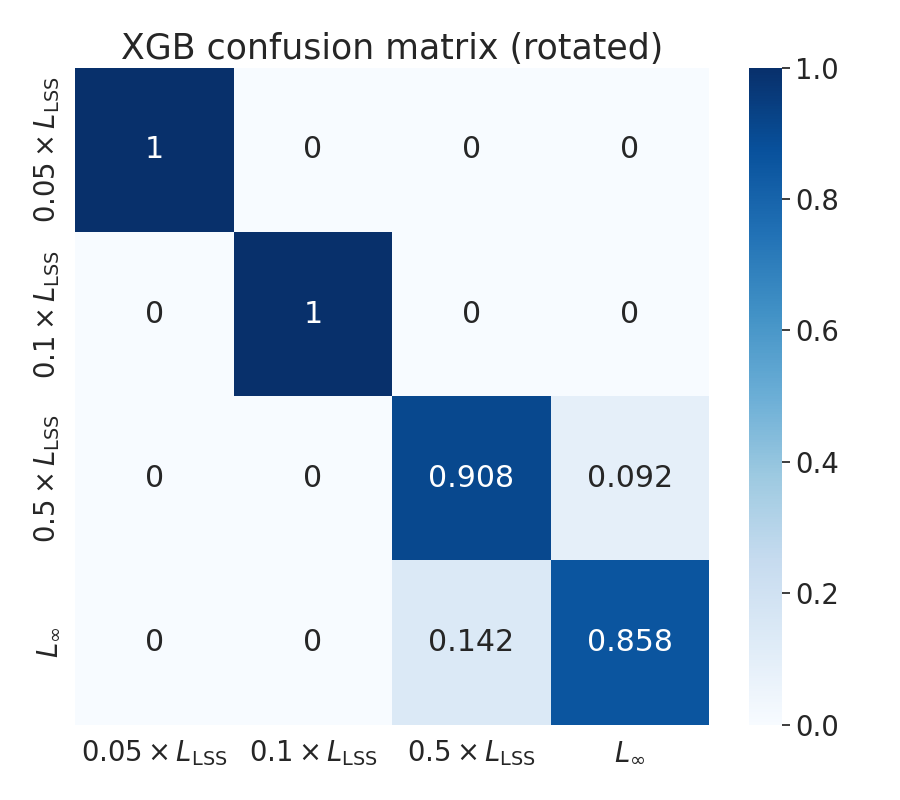}
        \label{fig_CM_RF_XGB:sub4}
    \end{subfigure}

    \caption{Normalized confusion matrices for the rotated and unrotated test datasets and for the random forest (RF) and \texttt{XGBoost} (XGB) algorithms. The left panels show the obtained classification accuracies for the unrotated test datasets, while the right panels show the accuracies for the randomly Wigner-rotated test data. In each confusion matrix the rows represent the true class, while the columns represent the predicted class. }
    \label{fig:CM_RF_XGB}
\end{figure}

The results obtained for decision-tree-based algorithms, i.e., random forests and \texttt{XGBoost}, indicate a classification accuracy of 99.8\%, when calculated on unseen unrotated test data. 
In the case where the training and the test datasets are randomly rotated, the combined test accuracy (over all dataset classes) decreases to $91\%-94\%$, depending on the algorithm (see \cref{table:random_forest_results}).
The performance on the studied topology classes is summarized in the confusion matrices presented in \cref{fig:CM_RF_XGB}.
The values along the diagonals list the accuracy for each topology class, while the off-diagonal values specify the percentage of
misclassified realizations.

\begin{table}[!h]
\centering
\begin{tabular}{lcc}
\hline \hline & \textbf{Random forests} & \textbf{\texttt{XGBoost}}  \\
\hline \textbf{Training (unrotated)} & 100\%  & 100\%  \\
\hline \textbf{Training (rotated)} & 100\%  & 100\% \\
\hline \textbf{Test (unrotated)} & 99.8\%  & 99.8\% \\
\hline \textbf{Test (rotated)} & 91.4\%  & 94.2\% \\ \hline \hline
\end{tabular}
\caption{Summary of the random forest and \texttt{XGBoost} results for different datasets. The percentages refer to the combined classification accuracy (over all the considered classes, i.e., $\{ 0.05, 0.1, 0.5, \infty  \} \times L_{\rm LSS}$). Unrotated and rotated refer to the two cases where the algorithms are trained and tested on unrotated and randomly rotated datasets, respectively. The accuracy for each individual class is listed in the confusion matrices of \cref{fig:CM_RF_XGB}. In each case, the algorithms are trained on data with $\ell \in [2,100]$.
}
\label{table:random_forest_results}
\end{table}

We find generally that for both datasets the two smallest classes $L=0.05 \times L_{\rm LSS}$ and $L = 0.1 \times L_{\rm LSS}$ are classified with 100\% accuracy.
For the randomly rotated realizations, both algorithms perform somewhat worse when distinguishing the $L = 0.5 \times L_{\rm LSS}$ and $L_{\infty}$ classes.
For these particular classes the \texttt{XGBoost} algorithm achieves significantly higher accuracies, i.e., $86\%-91\%$ of realizations are classified correctly, compared to $78\%-88\%$ for the random forest algorithm. 
When the realizations are misclassified, we generally find that the $L = 0.5 \times L_{\rm LSS}$ class is confused for the covering space realizations and vice versa.
This aligns with the expectation that smaller manifolds should be easier to classify, as they exhibit stronger correlations.

\begin{figure}
    \centering
    \begin{subfigure}{0.49\textwidth}
        \centering
        \includegraphics[width=1.0\textwidth]{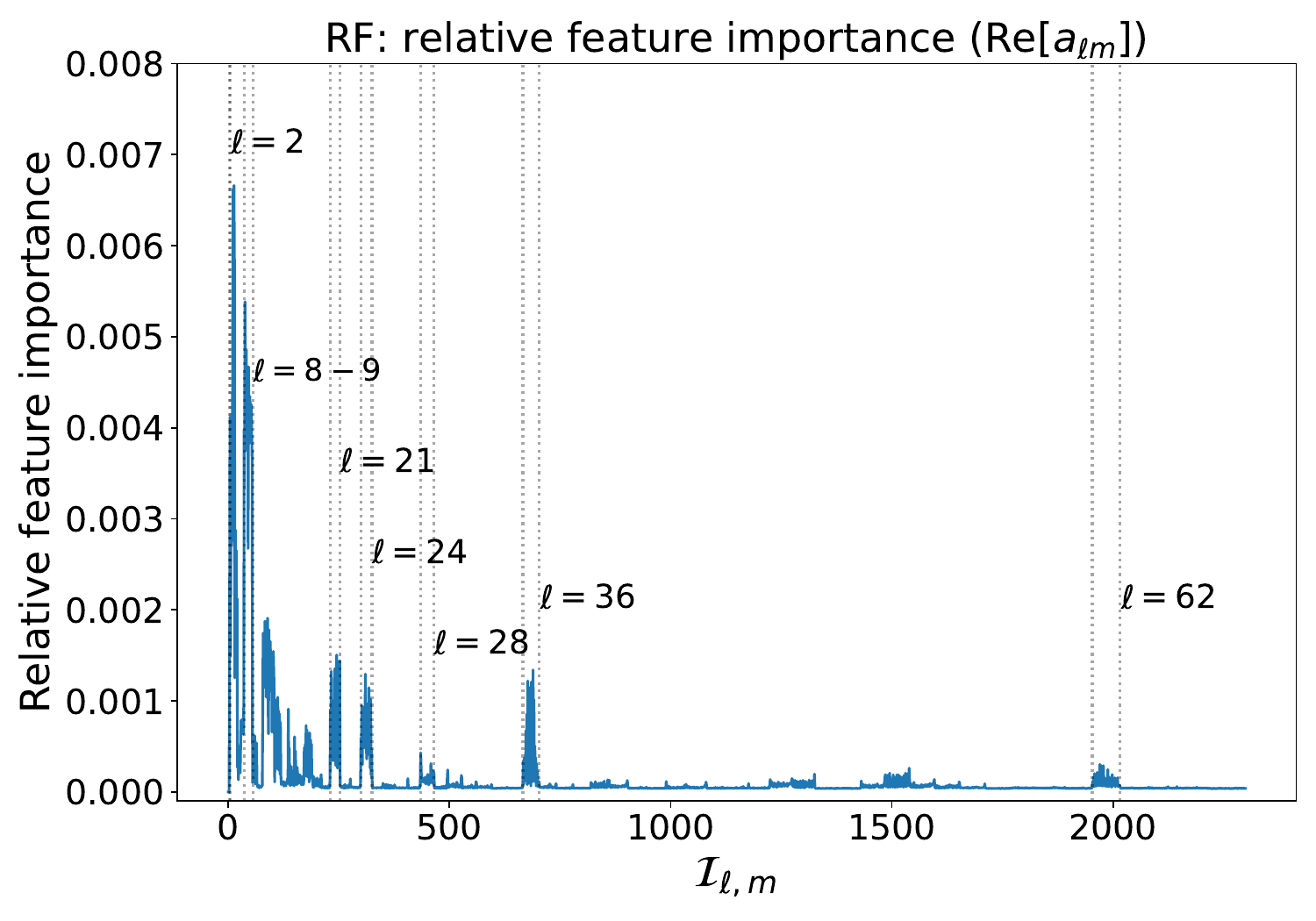}
        \label{fig:RF_feature_importance_1}
    \end{subfigure}
    \hfill
    \begin{subfigure}{0.49\textwidth}
        \centering
        \includegraphics[width=1.0\textwidth]{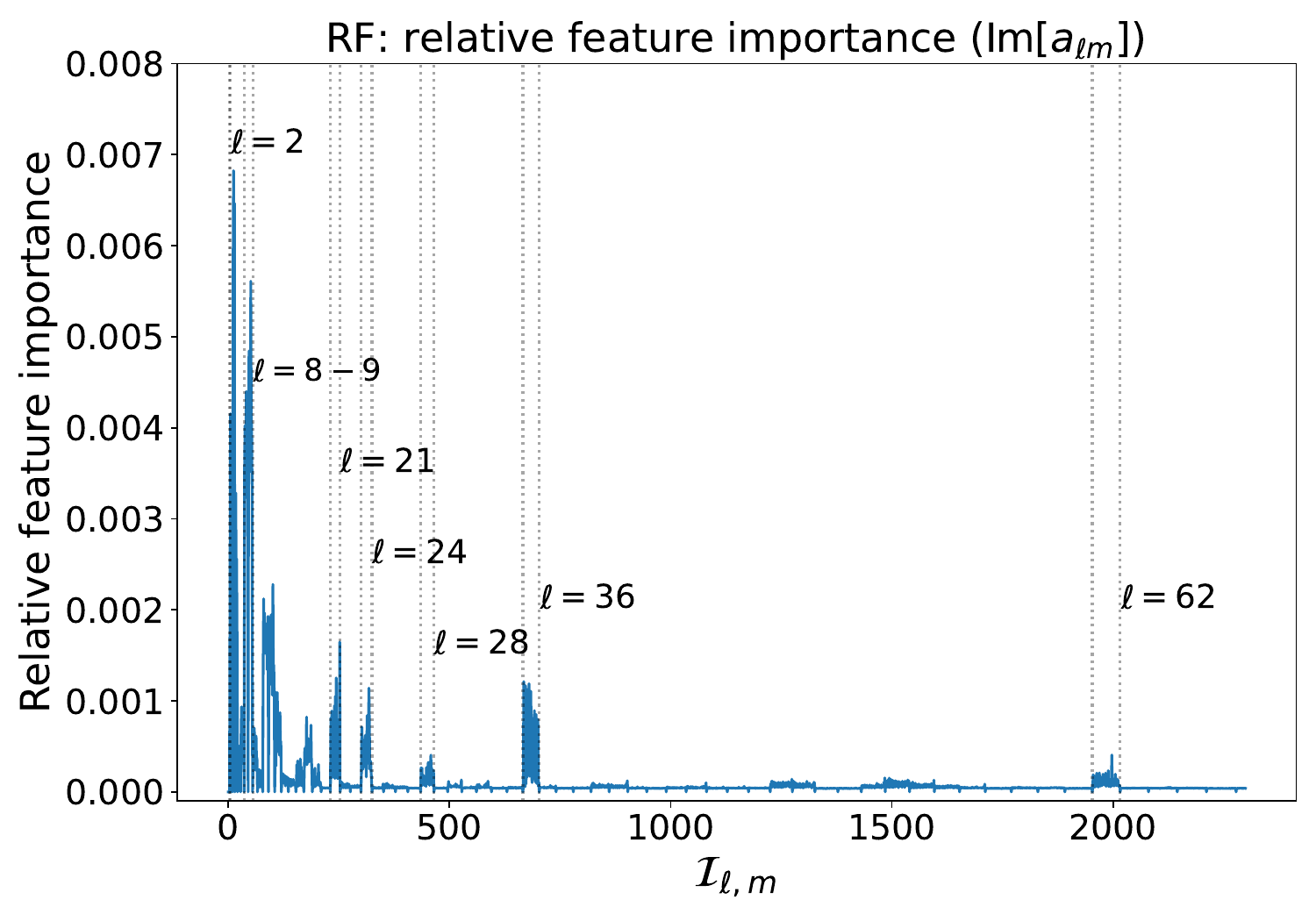}
        \label{fig:RF_feature_importance_2}
    \end{subfigure}
    \caption{Relative feature importance plots for the random forest (RF) algorithm. The feature importance for the real parts of the $a_{\ell m}$ are shown on the left and the imaginary parts are shown on the right. 
    The different spikes show the $a_{\ell m}$ coefficients that are particularly important for distinguishing the randomly rotated $\{ 0.05, 0.1, 0.5, \infty  \} \times L_{\rm LSS}$ topology classes. We have marked certain values of $\ell$ with dotted lines for reference. The feature importance for the $\vert\vert a_{\ell m}\vert\vert^2$ (the non-zero eigenvalue) that is also appended to the training data is not shown here. The $\vert\vert a_{\ell m}\vert\vert^2$ feature importance is equal to 0.04.}
    \label{fig:RF_feature_importance}
\end{figure}

\begin{figure}
    \centering
    \begin{subfigure}{0.49\textwidth}
        \centering
        \includegraphics[width=1.0\textwidth]{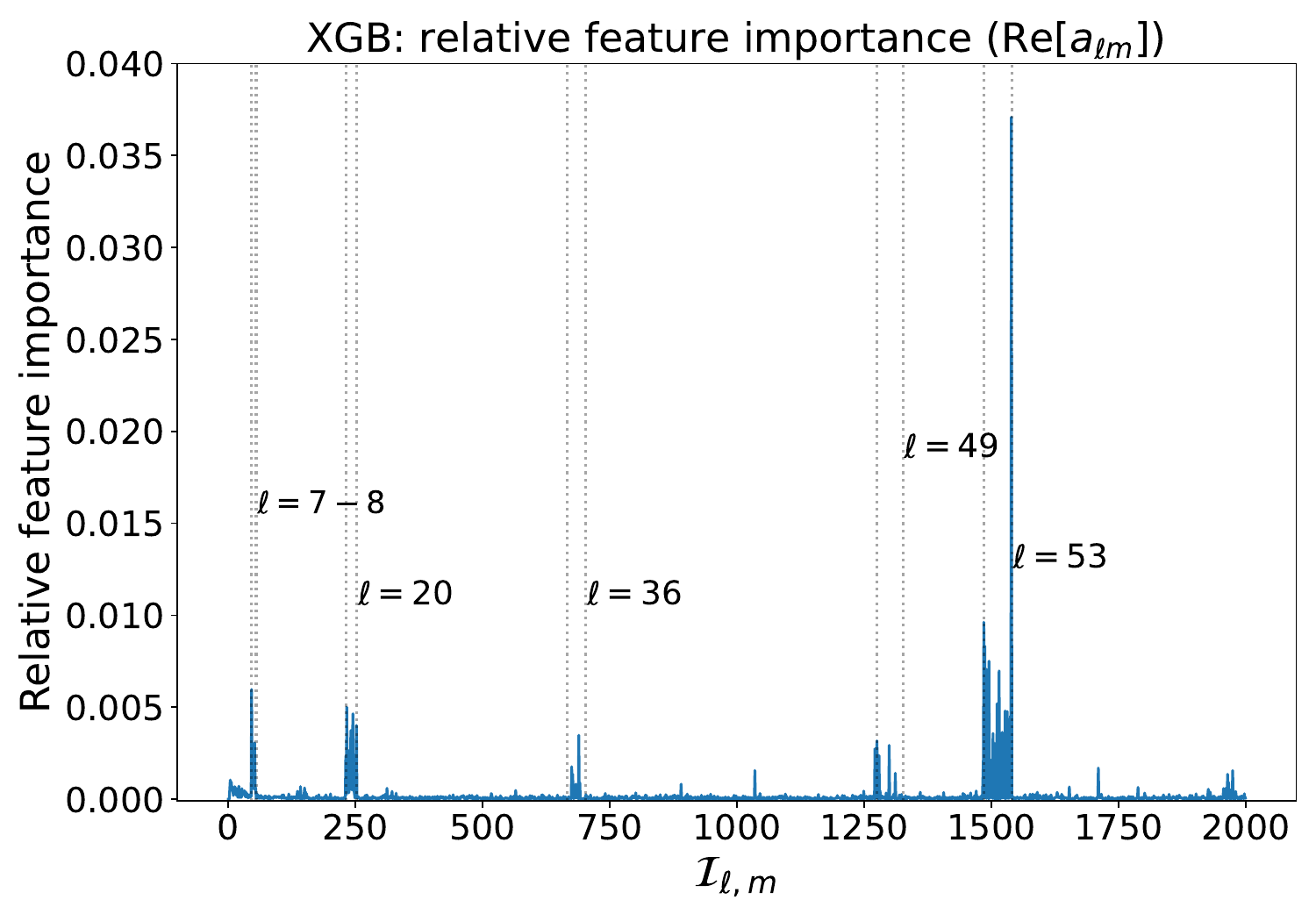}
        \label{fig:XGB_feature_importance_1}
    \end{subfigure}
    \hfill
    \begin{subfigure}{0.49\textwidth}
        \centering
        \includegraphics[width=1.0\textwidth]{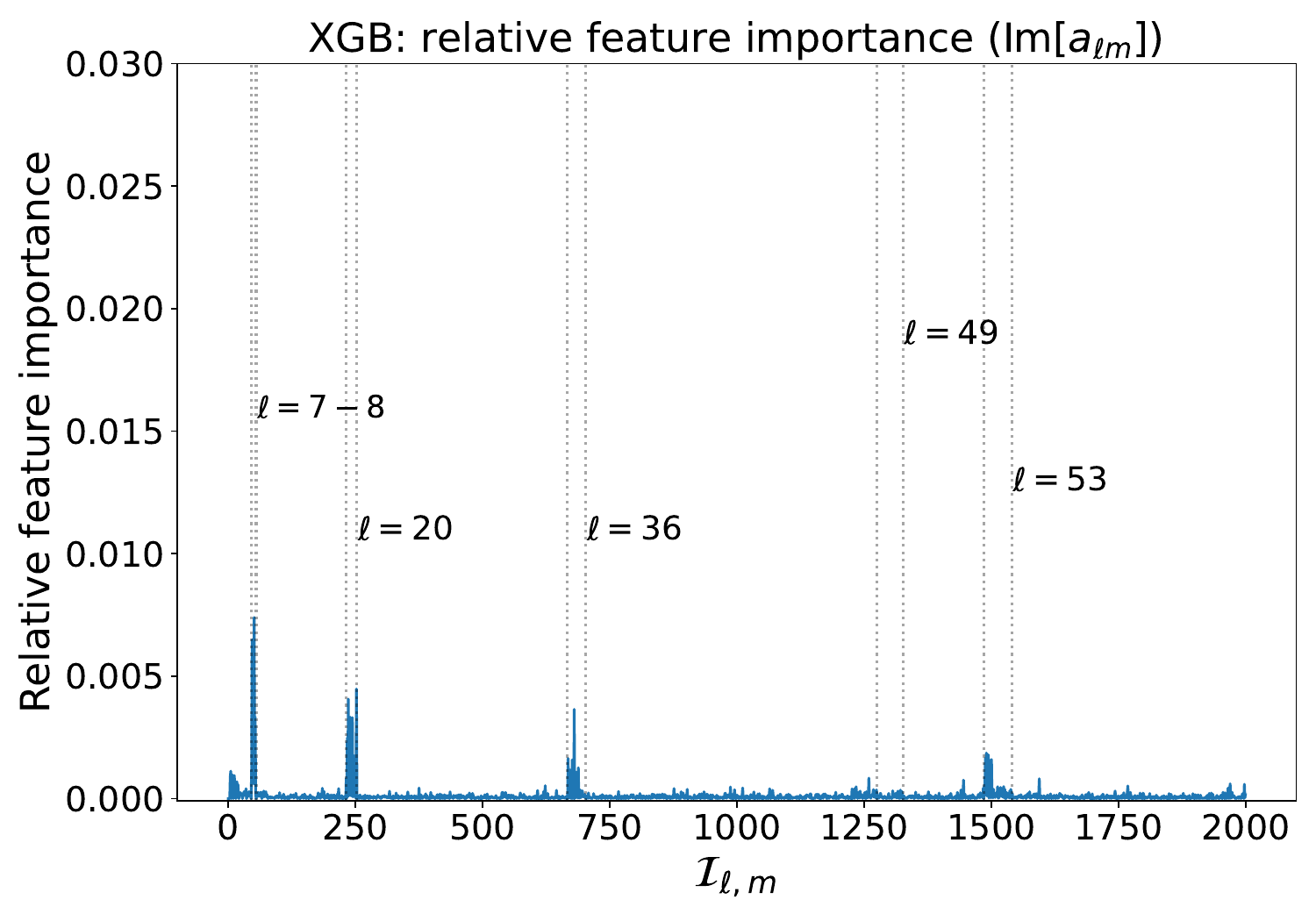}
        \label{fig:XGB_feature_importance_2}
    \end{subfigure}
    \caption{As in \cref{fig:RF_feature_importance}, but for the \texttt{XGBoost} (XGB) algorithm. The $\vert\vert a_{\ell m}\vert\vert^2$ feature importance (not shown here) has the value of 0.026.}
    \label{fig:XGB_feature_importance}
\end{figure}

One of the particular advantages of the decision-tree-based algorithms is the ability to calculate the relative feature importance. 
In our case, this specifically refers to the subset of $a_{\ell m}$ coefficients that are of particular importance for distinguishing the different topology classes. 
\cref{fig:RF_feature_importance} and \ref{fig:XGB_feature_importance} are feature importance plots for the random forest and \texttt{XGBoost} classification results.
The different spikes mark the  most important $a_{\ell m}$ coefficients.
The features are split into the real and imaginary values. 

For the random forests, we observe that the $\vert\vert a_{\ell m}\vert\vert^2$  feature (the non-zero eigenvalue) has the largest feature-importance score, likely due to being rotationally invariant.
We find that the $a_{\ell m}$ with small $\ell$ values are considerably more important than those with large $\ell$ values.
In the case of the \texttt{XGBoost} algorithm, however, the subsets of the $a_{\ell m}$ with $\ell \in \{7, 8, 20, 36, 49, 53 \}$ seem to be particularly important, comparable in importance to the $\vert\vert a_{\ell m}\vert\vert^2$ feature.
We also observe some minor differences between the real and imaginary features, but it is difficult to tell whether these point to the real features being more important that the imaginary ones, or  that the observed differences are due to inherent randomness of the algorithm.
Specifically, training the algorithms several times will result in the tallest peaks appearing in the same locations, but their relative importances will fluctuate due to dataset randomization and the inherent semi-randomness when choosing the features from which to build the decision trees.
Another minor difference between the real and imaginary features is that a significant subset of imaginary features have a feature importance of 0.
This specifically refers to the subset of $\{a_{\ell 0}\}$, which are purely real and hence their imaginary parts are not important for classification. 
Finally, in the random forest case, we also observe several important features corresponding to larger values of $\ell$, e.g., $\ell = 62$. 
The origin of the listed feature-importance spikes is  unclear, but may be related to the inherent symmetries of the $E_{1}$ topology.

\cref{table:NN_results} is a summary of the results for the neural-network-based algorithms for the rotated and unrotated datasets.
Specifically, we summarize the results for the 1D convolutional neural networks trained directly on the $a_{\ell m}$ data, as well as the 2D convolutional neural network results trained on the $\mathcal{C}_{\ell m \ell^{\prime} m^{\prime}}$ correlation data. 
The trend displayed in these results is generally the same as that shown for decision-tree-based algorithms -- we are able to accurately classify all the topology classes in the unrotated cases, while in the randomly rotated cases, classifying the largest $E_{1}$ realizations remains challenging.
When classifying unrotated realizations, both the 1D CNN and 2D CNN (\texttt{ResNet-50} and CVNN) algorithms obtain nearly perfect accuracy scores in the range of $99\%-100\%$ for both the test and the training datasets.
For the randomly rotated data, the algorithms obtain accuracies in the range  $91.8\%-93.7\%$, when calculated on unseen test data.
In this regard, the test results are slightly worse than those obtained by the \texttt{XGBoost} algorithm, however, any comparison between the algorithms should be done with caution, as different methodologies and different subsets of the dataset have been used in each case.
For example, we do not observe the CVNN model outperforming the 2D (real-valued) CNN and the \texttt{RestNet-50} model. 
Specifically, the CVNN model, in its current implementation, does not work with many of the commonly used layers employed by many of the real-valued networks described in this work (e.g., skip connections), making a direct comparison of the classification results difficult.

\begin{table}[h]
\centering
\begin{tabular}{lccc}
\hline \hline & \textbf{1D CNN} & \textbf{2D CNN} & \textbf{2D CVNN}  \\
\hline \textbf{Training (unrotated)} & 100\%  & 99.9\%  & 100\% \\
\hline \textbf{Training (rotated)} & 98.8\%  & 98.4\%  & 98.6\%\\
\hline \textbf{Test (unrotated)} & 99.8\%  & 99.4\% & 99.3\%\\
\hline \textbf{Test (rotated)} & 93.7\%  & 92.5\%  & 91.8\% \\  \hline \hline
\end{tabular}
\caption{As in \cref{table:random_forest_results}, but for the neural-network-based results. 1D CNN refers to the one-dimensional convolutional neural network trained directly on $a_{\ell m}$ data, 2D CNN refers to the \texttt{ResNet-50} architecture trained on two-dimensional $\mathcal{C}_{\ell m \ell^{\prime} m^{\prime}}$ data, and 2D CVNN refers to the two-dimensional complex convolutional architecture trained on $\mathcal{C}_{\ell m \ell^{\prime} m^{\prime}}$ data. The accuracy for each individual class is listed in the confusion matrices of \cref{fig:CM_results_NN}.}
\label{table:NN_results}
\end{table}

\begin{figure}
    \centering

    \begin{subfigure}{0.45\textwidth}
        \centering
            \includegraphics[width=0.95\linewidth]{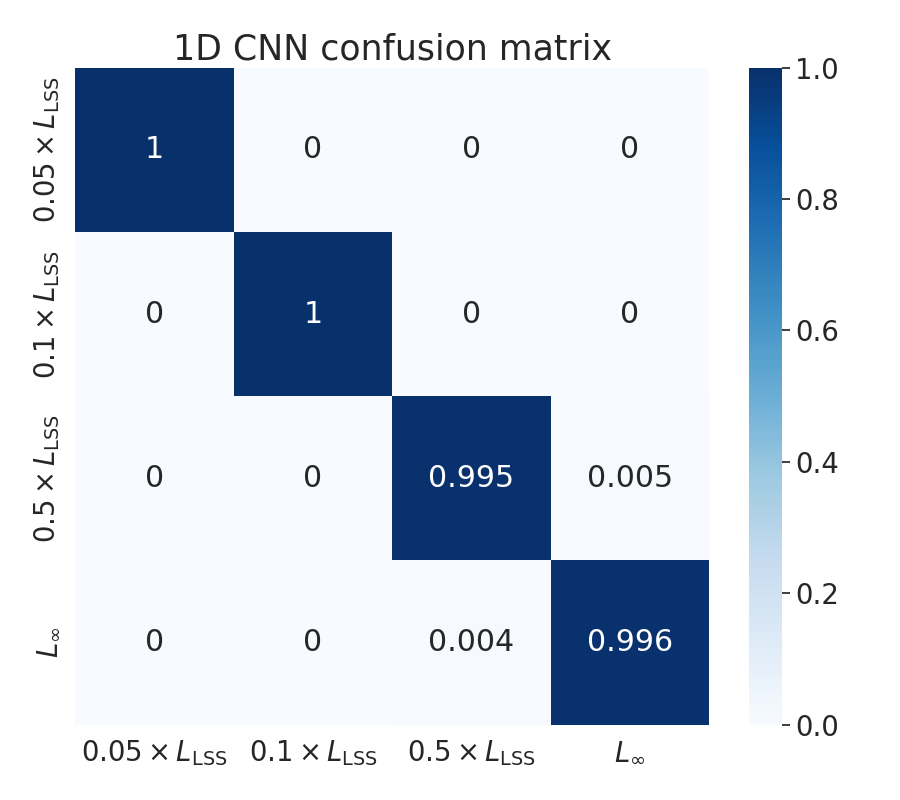}
        \label{fig:CM_results_NN:sub1}
    \end{subfigure}
    \begin{subfigure}{0.45\textwidth}
        \centering
        \includegraphics[width=0.95\linewidth]{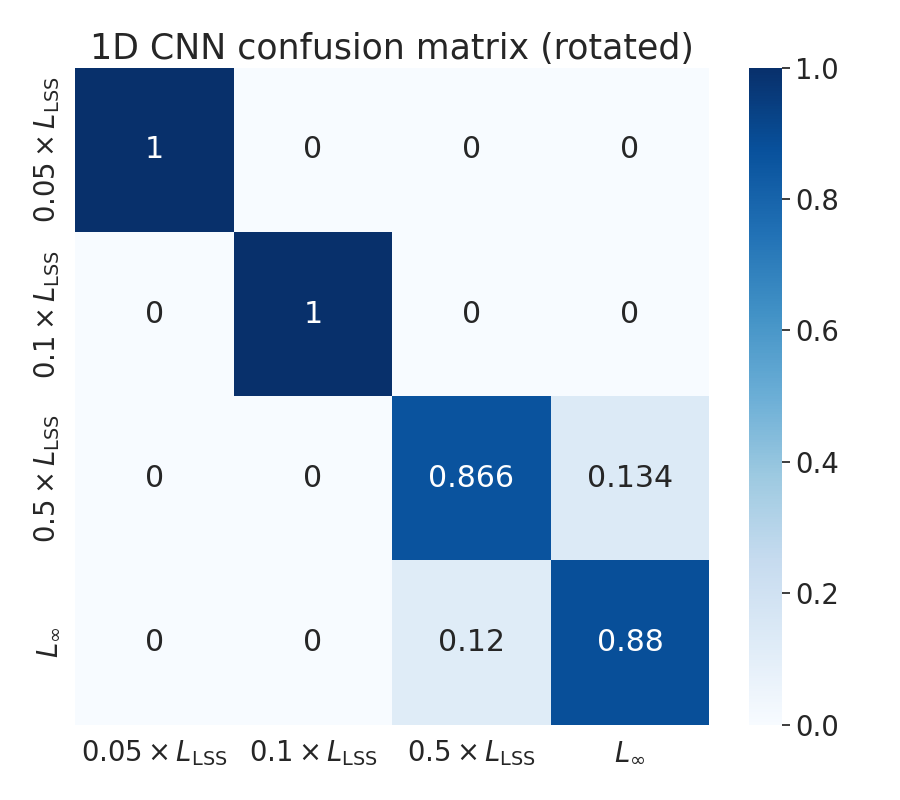}
        \label{fig:CM_results_NN:sub2}
    \end{subfigure}

    \medskip

    \begin{subfigure}{0.45\textwidth}
        \centering
        \includegraphics[width=0.95\linewidth]{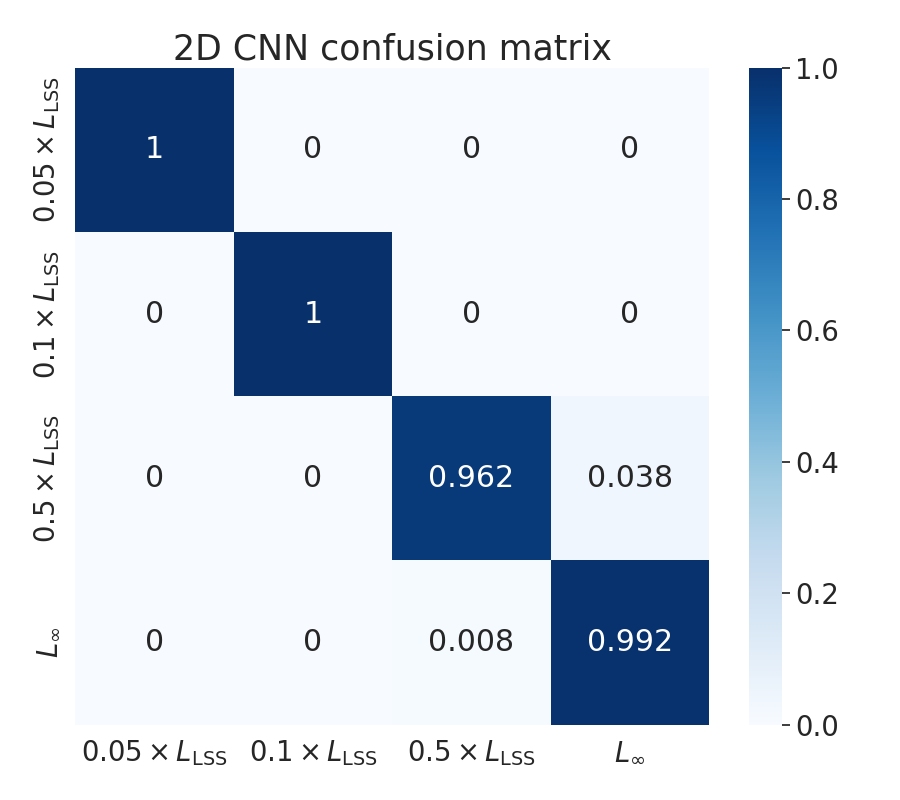}
        \label{fig:CM_results_NN:sub3}
    \end{subfigure}
    \begin{subfigure}{0.45\textwidth}
        \centering
        \includegraphics[width=0.95\linewidth]{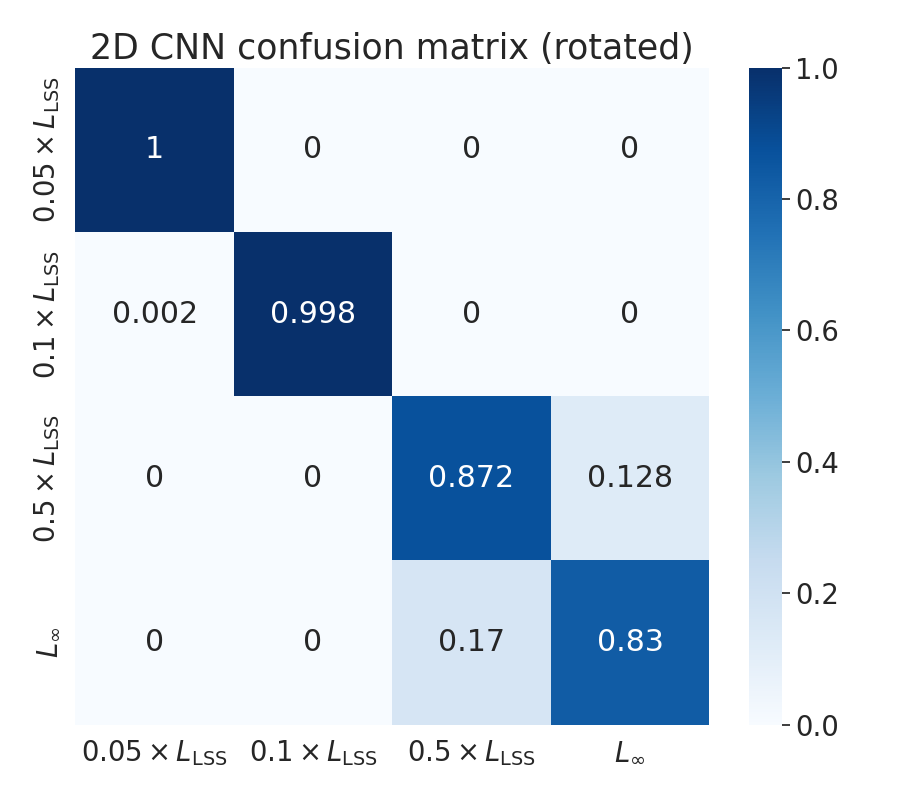}
        \label{fig:CM_results_NN:sub4}
    \end{subfigure}
    \medskip

    \begin{subfigure}{0.45\textwidth}
        \centering
        \includegraphics[width=0.95\linewidth]{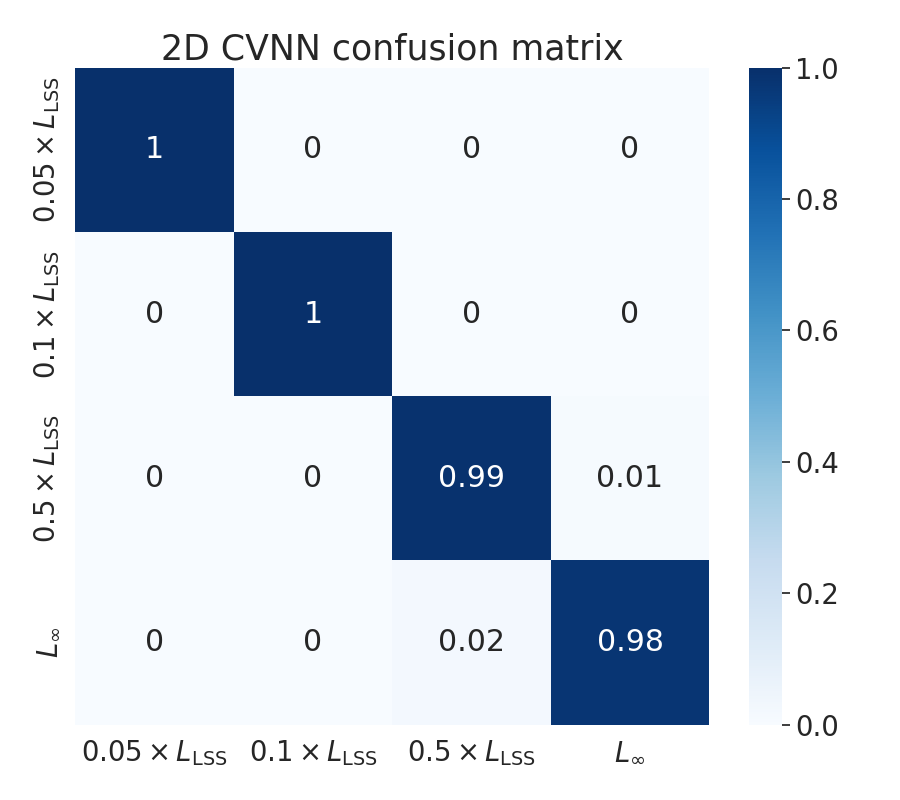}
        \label{fig:CM_results_NN:sub5}
    \end{subfigure}
    \begin{subfigure}{0.45\textwidth}
        \centering
        \includegraphics[width=0.95\linewidth]{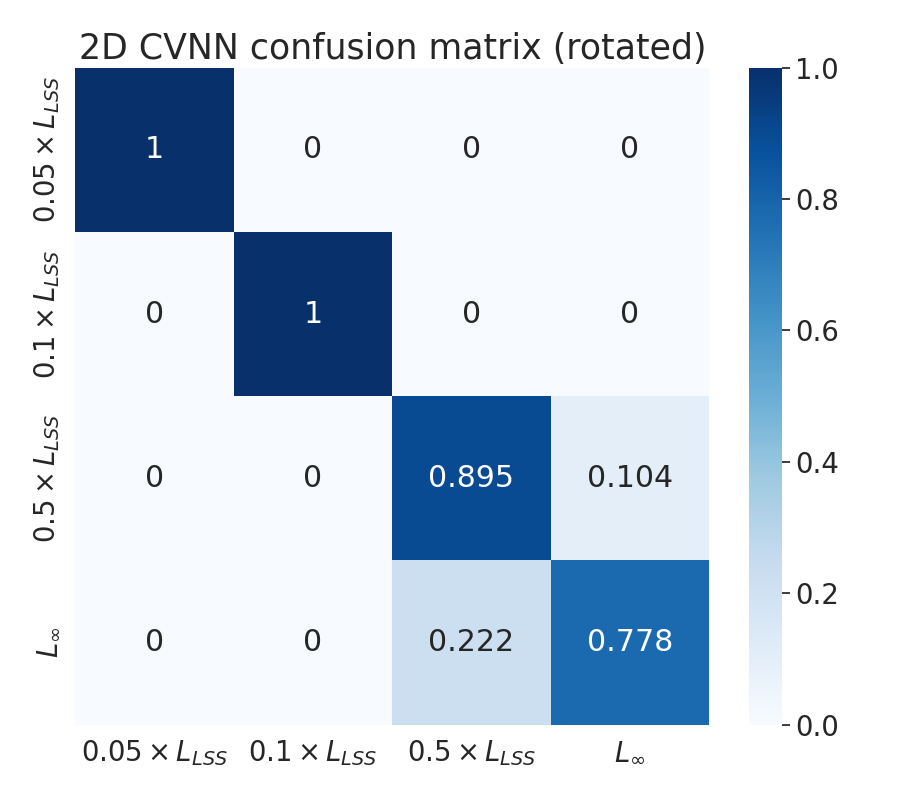}
        \label{fig:CM_results_NN:sub6}
    \end{subfigure}
    \caption{As in \cref{fig:CM_RF_XGB}, but for the neural-network-based algorithms 1D CNN, 2D CNN (\texttt{ResNet-50}), and CVNN.}
    \label{fig:CM_results_NN}
\end{figure}

As before, the confusion matrices for each dataset and each algorithm illustrate the individual topology class performance, as shown in Fig. \ref{fig:CM_results_NN}. Here we find that the algorithms classify the smaller $E_{1}$ topology classes, i.e., $0.05 \times L_{\rm LSS}$ and $0.1 \times L_{\rm LSS}$, nearly perfectly for all datasets (rotated and unrotated). 
When trained on randomly rotated realizations, the smaller classes are still classified with near-perfect accuracy, while classifying the larger ones, i.e., $0.5 \times L_{\rm LSS}$ and the covering space classes, is more challenging. 

It is also important to note that there are methods for calculating different feature importance statistics for CNNs (see, e.g., \rcite{Ahern2019, Wojtas2020, Lee2021}). This includes techniques such as activation maximization, layer-wise relevance propagation, and saliency maps. However, calculating these statistics is non-trivial and often requires a specific training procedure and sometimes specific layers incorporated in the architecture of the CNN. Similarly, while generally one can determine which input pixels were relatively more important, converting that to an interpretable statistic is difficult. For these reasons, we do not consider these methods in this work and leave it to be explored in future publications.

\subsection{$E_1$ realizations with $L \gtrsim L_{\rm LSS}$}
\label{subsection:results_L_L_LSS}

Previous work in the literature indicates that the KL divergence related to the signal of non-trivial topology in the CMB temperature anisotropies decreases with the size of the manifold. Specifically, there is a sharp decrease when the topology scale becomes larger than the diameter of the last-scattering surface, $L_{\rm LSS}$ (see Fig. 4 of \rcite{COMPACT:2022gbl}). 
In part this is related to the fact that we do not expect to see matched circle pairs in the CMB for manifolds of such size.
These matched circle pairs are composed of tightly correlated pixel pairs. 
When $L>L_\mathrm{LSS}$, these tight pixel-pixel correlations are absent, and we may expect a qualitative shift in the nature of the correlations -- certainly the KL divergence between a compact manifold and the covering space declines.
Thus, a pressing question is whether we can use the same machine learning methods that are described in this work in order to classify realizations of $E_{1}$ larger than $L_{\rm LSS}$.
Here, as a test, we present the results for a random forest classifier trained on $a_{\ell m}$ realizations falling into two classes: $E_1$ topology with $L  = 1.01 \times L_{\rm LSS}$ and the covering space.
The results are shown for different dataset sizes ranging between 200 and 200,000 realizations, generated via the Cholesky decomposition method (see \cref{appendix:large_L_results} for technical details). 

\begin{figure}
  \centering
  \includegraphics[width=0.65\textwidth]{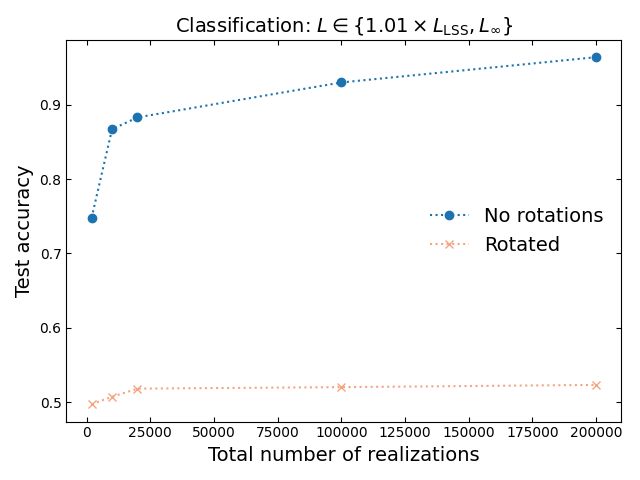}
  \caption{Classification results for $E_{1}$ realizations with $L = 1.01 \times L_{\rm LSS}$ versus the covering space realizations. All realizations are generated with $\ell_{max} = 30$ using the Cholesky decomposition method. The test dataset accuracy is calculated on unseen data, which is equal in size to the 20\% of the total dataset size.}
  \label{fig:large_L_results}
\end{figure}

The results are summarized in \cref{fig:large_L_results}. 
For unrotated realizations, the performance of the random forest classifier depends significantly on the total dataset size.
In order to obtain accuracies of $\sim 90$\%, around 30,000 realizations are needed.
For a dataset of 200,000 realizations the algorithm obtains a test dataset accuracy of 98\%, i.e., comparable to that obtained for realizations with $L < L_{\rm LSS}$.
This is an encouraging result as it shows that, even for large unrotated $E_{1}$ realizations, the classification accuracy is limited primarily by the dataset size.
If sufficient number of realizations is available, high classification accuracies can be obtained.

For rotated realizations, however, the results are less optimistic. 
In particular, increasing the dataset size from 20,000 realizations to 200,000 realizations has almost no effect on the test classification accuracy,
which is a dismal  $52$\%. 
Note that a similar decrease in accuracy when training on the randomly rotated dataset (albeit to a much lesser extent) is observed for the results in \cref{subsection:results_small_E1}.
These results likely indicate that the issues related to rotated realizations cannot simply be resolved by adding more data. 
New techniques are required.

\section{Discussion and conclusions}
\label{section:discussion}

In this work, we have explored the effectiveness of two categories of machine learning algorithms, decision-tree-based random forests and \texttt{XGBoost} classifier, along with 1D and 2D convolutional neural networks, for classification of harmonic-space realizations of CMB full-sky temperature maps for classes of cosmic topology. We have tested the outlined algorithms on the specific case of cubic $E_{1}$ (or 3-torus) topology of four different size classes, $\{ 0.05, 0.1, 0.5, \infty\} \times L_{\rm LSS}$. In addition, we have assessed the feasibility of distinguishing $a_{\ell m}$ realizations of cubic $E_{1}$ with the size scales $L \gtrsim L_{\rm LSS}$ from the  $a_{\ell m}$ realizations of the covering space. We have generated two sets of results in each case: one for the case where the orientation of each realization is aligned with the coordinate frame of the cubic $E_{1}$ manifold and one where each realization is randomly rotated. Even though our focus in this work has been on manifolds with small size scales, which are excluded by current observational data through, e.g., circles-in-the-sky searches, the obtained results identify the prospects and the main challenges for developing machine learning methods that are capable of accurately classifying observationally viable topologies.

All the machine learning algorithms we have tested can distinguish realizations of $E_{1}$ topology with a cubic fundamental domain (and of the covering space) with high accuracy when the realizations in the dataset are unrotated relative to a coordinate system aligned with the edges of the cube.
Although we have not presented these,
the same conclusion applies when all the realizations in all the data sets are rotated by the exact same arbitrary rotation.
In other words, if the coordinate system of the observer is oriented relative to the coordinate axes in the same exact way for each and every realization, the nearly perfect results observed for the unrotated realization are recovered.

For datasets where each realization is randomly rotated relative to one another, the classification results are generally worse for the $L = 0.5 \times L_{\rm LSS}$ and the covering space classes. 
To some extent, this is not surprising, as Wigner rotations scramble the $m$'s for each $a_{\ell m}$ in the dataset, 
and do so in a different way for each $\ell$,
making it difficult for the algorithms to learn the features encoded in the correlations between the different $m$ and $\ell$ values.
Conceptually, this problem can be compared to classic CNN architectures being generally bad at training on rotated image data. 
In fact, a more accurate analogy would be training a CNN on a set of images, where the pixels of each image have been randomly shuffled. More generally, standard CNN architectures do not allow for learning rotationally invariant and equivariant features.
There are, however, custom CNN architectures that are capable of learning rotationally invariant features, which does allow dealing with rotated image data more easily and accurately \cite{Marcos2016, Kim2020, Alt2021, Hanlin2022}.
Hence, one possible direction of future work is exploring such more complex neural network architectures.
Given that we can obtain an accuracy greater than $\sim 90$\% for realizations with $L < 0.5 \times L_{\rm LSS}$ without imposing rotational invariance, it is reasonable to expect that using one of the rotationally invariant architectures mentioned above would only further increase the classification accuracy. 

The importance of the orientation of the coordinate system when looking for signal of non-trivial topology has been previously considered in \rcite{Kunz:2006}.
Specifically, the authors concluded that the likelihood of detecting a signal of non-trivial topology is maximized when the orientation of the harmonic-space realizations is aligned with the orientation of the covariance matrix that is used for comparison (Fig.~1 in \rcite{Kunz:2006}).
This result, in particular, offers some hints on how an analytic covariance matrix for a specific topology class could be used to align a given randomly rotated $a_{\ell m}$ realization. 
Specifically, one could calculate rotation curves akin to those shown in Fig.~1 of \rcite{Kunz:2006}, and then use the maximum value of the correlation coefficient or the likelihood to further rotate a given realization to one of the optimal orientations.
Alternatively, such rotation curves could be used as an extra input to our algorithms. Based on the findings of \rcite{Kunz:2006}, such an approach is promising; however, it is beyond the scope of this  work and is left for future publications.

Our investigation indicates that machine learning techniques described in this work are also capable of correctly classifying harmonic-space realizations for non-trivial topology manifolds of $L \gtrsim L_{\rm LSS}$. 
Specifically, in \cref{subsection:results_L_L_LSS} we have calculated the classification accuracy for significantly larger datasets generated using a different method (Cholesky decomposition).
The obtained results indicate that random forests trained on a large dataset with $\sim 10^{5}$ harmonic-space realizations can be classified with accuracies comparable to those presented in \cref{table:random_forest_results}.
Increasing the size of the training dataset  results in a continuous increase of the accuracy scores on the test data. However, that is only the case when trained on unrotated harmonic-space realization data.
Increasing the total dataset size seems to have almost no effect when training on randomly rotated realizations.
This shows that the issues related to training on rotated realizations cannot be solved simply with a  larger training dataset.
Nonetheless, if these problems can be addressed, our results indicate that machine learning offers a set of powerful techniques to detect the signal of non-trivial topology.

Note that in this work we have only explored a small region of the parameter space of the $E_{1}$ topology, i.e., cubic manifolds with $L < L_{\rm LSS}$ and $L \gtrsim L_{\rm LSS}$. In future work this will be extended to non-cubic and possibly tilted (see, e.g., Section 3.1 of \rcite{COMPACT:2023_paper2A}) manifolds.
Previous work also indicates that different classes of topology can have different KL divergence values for the same topology scale (e.g., see Fig.\ 4 of \rcite{COMPACT:2022gbl}).
What is more, even if a given topology has the same value of KL divergence, it does not necessarily mean that the features encoded in harmonic-space realizations will be encoded in the same way. 
Hence, another natural extension of this work is to explore a larger set of topologies.
Cubic 3-torus topology has significant accidental rotational symmetries (i.e., the symmetry of a cube), whereas generic topologically non-trivial manifolds lack those symmetries.
 It  therefore seems likely that more general $E_1$ and  other topologies  will be easier to classify. Our preliminary tests indicate that the classification accuracies  for $E_{2}$-$E_{6}$ are generally larger than those obtained in this work.
However, these results require further analysis, which is left for future publications.

Furthermore, we  note that a spherical harmonic expansion is not the only possible representation of the CMB sky.
An alternative representation, the multipole vector formalism \cite{Copi:2003kt},  has been used extensively in studies of the CMB anomalies \cite{Copi:2003kt, Bielewicz:2008ga,Oliveira:2018sef}.
A distinguishing feature of multipole vectors is that rotations of the coordinate system correspond to a rigid rotation of the set of multipole vectors.
It is therefore easy to define rotationally invariant quantities, such as dot products of multipole vectors.
These features of the multipole vector formalism warrant further investigation using machine learning, which is left for future work. 

Techniques presented in this work deal specifically with the harmonic realizations of the full sky maps. Looking forward, an interesting question is whether techniques like this could be applied to real observational data. 
This raises a number of important questions, such as what would be the effects of working with the harmonic realizations of partial or masked sky maps. 
Presumably, effects such as these would break isotropy and introduce extra harmonic space correlations, which would be challenging to distinguish from the correlations induced by non-trivial topology. 
Similarly, effects such as foreground residuals and beam asymmetries would have to be accounted for. While studying these systematics is beyond the scope of this work, it is clear that any successful method applied to real CMB data would have to take into account these observational effects. Hence, correctly accounting for such effects is a natural direction for future work.   

The problem of distinguishing harmonic space realizations in non-trivial topology from the corresponding covering space realizations has previously been explored using Bayesian and frequentist approaches in \rcite{Kunz:2006}. 
While we did not attempt to use likelihood-based methods in this work, the techniques described in \rcite{Kunz:2006} offer a valuable point of comparison. 
Namely, the authors of \rcite{Kunz:2006} have calculated the likelihood function using $10^{4}$ simulated harmonic space realizations of $E_{1}$ topology of different sizes. 
The likelihood calculations have then been used to distinguish between two cases of special interest: one of an infinite universe (covering space) and the other corresponding to a universe with an $E_{1}$ topology characterized by the relevant correlation matrix. 
The statistical significance with which the two scenarios can be distinguished has been quantified by the signal-to-noise ratio statistic (Eq. 12 in \rcite{Kunz:2006}). 
For instance, a universe with $L \approx 0.66 \times L_{\rm LSS}$ (or $T[4,4,4]$ in the units of \rcite{Kunz:2006}) can be distinguished from the covering space with $5\sigma-9.3 \sigma$ significance depending on the value of $\ell_{\rm max}$ and the details of the normalization. 
In our case, for instance with the 1D CNN algorithm, we can distinguish the $L = 0.5 \times L_{\rm LSS}$ class from the covering space with $86.6\% - 88\%$ accuracy for the rotated datasets and $99.5\% - 99.6\%$ accuracy for the non-rotated datasets. 
However, it is important to note that the correlation matrices used in \rcite{Kunz:2006} do not include the integrated Sachs-Wolfe effect (ISW) contribution. 
Properly accounting for the ISW effect should in principle significantly lower the detection power of the outlined likelihood approach, as the authors have pointed out in their section VIII C. 
In our work, the used correlation matrices do include the ISW and the Doppler shift contributions.

Nonetheless, the results quoted above indicate that likelihood-based methods can generally distinguish $E_{1}$ from the covering space with statistical significance higher than that of the machine learning methods described in this work. 
However, it is important to point out the limitations of the likelihood-based approaches.
Firstly, the obtained results are very sensitive to the relative orientation of a given $\{ a_{\ell m} \}$ set and the corresponding correlation matrix used in the likelihood calculation (this is fundamentally the same issue of dealing with rotated realizations that we have described in this work).
The likelihood generally needs to be calculated for over $10^{6}$ orientations to obtain the results quoted in \rcite{Kunz:2006}. 
Furthermore, a full likelihood search would require considering 17 non-trivial topologies, each with up to 6 real parameters specifying the manifold, up to 3 parameters specifying the location of the observer, and up to 3 further parameters specifying the orientation. 
Machine learning methods offer a computationally cheaper and more efficient  alternative: after a training stage, which generally takes several hours, predictions can be generated for thousands of realizations (and over thousands of orientations) in a matter of seconds.

Another important aspect to consider is that in the likelihood-based approaches it is difficult to know what particular features in harmonic space (or pixel space) carry the information that allows distinguishing non-trivial topology from the covering space. 
In other words, given a CMB map of non-trivial topology, what information specifically distinguishes it from a corresponding covering space map? 
Machine learning approaches offer a number of techniques to extract this information. 
As illustrated in our work, the feature importance analysis provides information on which multipoles (or which angular scales in the sky maps) carry the information that is particularly useful for distinguishing that given realization from the covering space realizations. 
Such information is particularly valuable when considering $a_{\ell m}$ realizations with $L > L_{\rm LSS}$, as in this regime there are no matched-circle pairs, and hence, it is not trivial to deduce how the information that allows distinguishing between different topologies is encoded in pixel space. 
In future work we hope to combine the outlined feature importance analysis with other methods such as convolutional neural network layer-wise relevance propagation or saliency analysis. 
Our hope is to also use autoencoders and information-maximizing neural networks (IMNN's) \cite{Charnock2018} to extract features related to non-trivial topology and do likelihood-free inference \cite{Alsing:2018eau}. 

In summary, we have demonstrated that machine learning techniques are promising for using harmonic space realizations to distinguish between classes of topological manifolds, including distinguishing non-trivial topology from the covering space. Machine learning methods presented in this work also offer a computationally efficient way of extracting features related to non-trivial topology.
Nonetheless, our freedom to rotate our coordinate system is the current principal challenge.

\acknowledgments
We thank Roger French for valuable conversations. This work has made use of the High Performance Computing Resource in the Core Facility for Advanced Research Computing at Case Western Reserve University. This work has also made use of the High Performance Computing Resources available as a part of the Pitzer Cluster provided by the Ohio Supercomputer Center. 

A.T.\ is supported by the Richard S.\ Morrison Fellowship. F.C.G\ is supported by the Presidential Society of STEM Postdoctoral Fellowship at Case Western Reserve University.
Y.A.\ acknowledges support by the Spanish Research Agency (Agencia Estatal de Investigaci\'on)'s grant RYC2020-030193-I/AEI/10.13039/501100011033, by the European Social Fund (Fondo Social Europeo) through the  Ram\'{o}n y Cajal program within the State Plan for Scientific and Technical Research and Innovation (Plan Estatal de Investigaci\'on Cient\'ifica y T\'ecnica y de Innovaci\'on) 2017-2020, and by the Spanish Research Agency through the grant IFT Centro de Excelencia Severo Ochoa No CEX2020-001007-S funded by MCIN/AEI/10.13039/501100011033. J.R.E.\ acknowledges support from the European Research Council under the Horizon 2020 Research and Innovation Programme (Grant agreement No.~819478). T.S.P acknowledges financial support from the Brazilian National Council for Scientific and Technological Development (CNPq) under grants 312869/2021-5 and 88881.709790/2022-01.
C.J.C., A.K., D.P.M., and G.D.S.\ acknowledge partial support from NASA ATP grant RES240737; G.D.S.\ from DOE grant DESC0009946; Y.A., O.G., G.D.S., S.S., and Q.T.\ from the Simons Foundation; Y.A., A.H.J., and G.D.S.\ from the Royal Society (UK); and A.H.J.\ from STFC in the UK\@.

\appendix
\section{Training procedures and settings used for machine learning algorithms}
\label{appendix:A}

\subsection{Random forests and extreme gradient boosting classifier}
\label{appendix:random_forests}

Our analysis shows that the most important parameters for the random forest classifier are the number of estimators (\texttt{n\_estimators}) and the maximum depth of the trees (\texttt{max\_depth}).
After performing a grid search on a small sample dataset to determine the optimal set of these parameters, we find that the number of estimators should be set to around $30\%-35\%$ of the number of features, i.e., $\texttt{n\_estimators} \approx 3000-3500$ for a full dataset with $\ell_{\rm max} = 100$.
Similarly, not adding a limit to the maximum tree depth results in the highest test dataset accuracies. 

For the \texttt{XGBoost} classifier, we use the same exact dataset with the same data ordering.
Similarly, we find that the key parameters for obtaining high classification accuracy are the number of boosted trees (\texttt{n\_estimators}), maximum three depth for the base learners (\texttt{max\_depth}), and the boosting learning rate parameter (\texttt{learning\_rate}). After performing a grid search for the optimal number of boosted decision trees, we find the optimal value to be 2000.

Here we summarize the parameter values used by both classifiers:
\begin{enumerate}[itemsep=0.05pt]
\item \textbf{Random forest classifier:} 
\begin{itemize}[itemsep=0.1pt]
    \item Number of tree estimators: \texttt{n\_estimators = 3500}
    \item Maximum tree depth: \texttt{max\_depth = None}
    \item Function to measure the quality of splits: \texttt{gini}
    \item Number of features to consider when looking for optimal splits: \texttt{auto}
    \item All the other settings are set to their default values listed in \rcite{sklearn_random_forests}.
\end{itemize}
\item \textbf{\texttt{XGBoost} classifier:}
\begin{itemize}[itemsep=0.1pt]
    \item Number of tree estimators: \texttt{n\_estimators = 2000}
    \item Maximum tree depth: \texttt{None}
    \item Boosting learning rate parameter: \texttt{learning\_rate = 0.1}
    \item Type of the used booster: \texttt{booster = gbtree}
    \item Feature importance calculation setting: \texttt{importance\_type = gain}
    \item All the other settings are set to their default values listed in \rcite{xgboost_documentation}.
\end{itemize}
\end{enumerate}

\subsection{1D convolutional neural networks}
\label{appendix:1D_CNN_architecture}

While the results of the decision-tree-based algorithms do not depend on the $a_{\ell m}$ realization data ordering, this is not generally the case for CNNs.
To investigate this for each dataset, we test the two $(m, \ell)$ and $(\ell, m)$ orderings of $a_{\ell m}$ values.
To asses these orderings we run the 1D CNN training procedure for different subsets of the total dataset with $\ell_{\rm max} = \{10, 20,50,100\}$.
In each case, the classification accuracies are then compared for the two different orderings.
We generally find that the ordering does not have a significant effect on the classification accuracy, and hence, we choose the $(\ell, m)$ ordering, which is easier to interpret.  

Similarly, we test different ways of representing real and imaginary parts of the harmonic-space realizations.
Namely, we test the representation of the real and imaginary parts as different neural network channels rather than simply appending the imaginary part to the real values. However, we do not find it to have a significant effect on the classification accuracy.
A similar question is that of the ordering of the $a_{\ell m}$ real and imaginary parts, i.e., one can simply append the array of the real parts to that of the imaginary parts for each realization, resulting in: $[\{ a_{\ell m}^{\rm Re} \}, \{ a_{\ell m}^{\rm Im} \}]$.
Alternatively, one can consider appending the two components for each $\ell$, resulting in $[ a_{00}^{\rm Re}, a_{00}^{\rm Im}, a_{10}^{\rm Re}, a_{10}^{\rm Im}, \ldots]$.
Here we find that the change in the classification accuracy is not generally larger than the usual fluctuations in the accuracy due to the inherent randomness of the training procedure (e.g., randomly choosing a subset of the training/test data, randomizing the initial weights of the network, \textit{etc}).
We therefore choose to work with the simple approach of appending the real values to the imaginary values for the full range of $\ell$'s rather than for each $\ell$ in the dataset. 

One key difference we find is that 1D CNNs are much more sensitive to the number of features in the training dataset. In other words, we generally find that including only a subset of the available $\ell$ values, e.g., $\ell \in [2,\ell_{\rm max} = 50]$, results in higher test dataset accuracy score.
Particularly, we repeat the training procedure for different values of $\ell_{\rm max} \in \{10,20,30,50,100 \}$ and find that values of $\ell \leq \ell_{\rm max} = 50$ result in a similar test dataset accuracy, with $\ell_{\rm max} = 50$ giving the highest value. On the contrary, $\ell_{\rm max} = 100$ results in worse performance, likely due to feature over-abundance compared to the size of the available dataset. 

For the model architecture, we choose to work with multiple 1D convolutional layers with a varying number of filters of different sizes.
This is motivated by the fact that we expect the different pairs of multipoles to be correlated in non-trivial topologies (e.g., the \textit{checkerboard} pattern in \cref{fig:cm_analytic_plots}). 
Given the chosen data ordering, this implies that we expect non-local correlations between the different features in our dataset, and having multiple convolutional layers with different kernel sizes is a known method for extracting non-local features and has been considered in architectures used for audio and music genre classification, electroencephalography (EEG) classification, and transiting exoplanet signal detection \cite{Allamy2021, Kiranyaz2021, Mattioli2021, Iglesias2023}.
The convolutional layers are then appended by 1D max pooling and dropout layers in order to assist the feature extraction and to reduce overfitting.
The final layers in the architecture are the four dense layers with \texttt{LeakyRelu} activation functions.
The architecture is summarized in \cref{table:1D_CNN_architecture}.

\begin{table}[t!]
\centering
\begin{tabular}{lccc}
\hline \hline & \textbf{Activation} & \textbf{Output shape} & \textbf{Parameters} \\
\hline \hline Input map & - & (None, 2652, 32) & 128 \\
\hline \texttt{Conv1D} & \texttt{LReLU} & (None, 2652, 64) & 10304 \\
\hline \texttt{Conv1D}  & \texttt{LReLU} & (None, 2652, 128) & 57472 \\
\hline \texttt{Conv1D} & \texttt{LReLU} & (None, 2652, 256) & $295 \mathrm{~K}$ \\
\texttt{Max Pooling 1D} & - & (None, 1326, 256) & - \\
\texttt{Dropout} & - & (None, 1326, 256) & - \\
\texttt{Flatten} & - & (None, 339456) & - \\
\hline \texttt{Dense} & \texttt{LReLU} & (None, 512) & $173 \mathrm{~M}$ \\
\hline \texttt{Dense} & \texttt{LReLU} & (None, 256) & $131 \mathrm{~K}$ \\
\hline \texttt{Dense} & \texttt{LReLU} & (None, 128) & $32 \mathrm{~K}$ \\
\hline \texttt{Output layer} & \texttt{Softmax} & 4 & 516 \\
\hline \hline \multicolumn{3}{l}{ \textbf{Total trainable parameters: }} & $\mathbf{174 \; M}$ \\
\hline \hline
\end{tabular}
\caption{Summary of the 1D CNN architecture used to train on the $a_{\ell m}$ data. The model is compiled with the \texttt{Adam} optimizer with the learning rate of $10^{-5}$ and \textit{sparse categorical crossentropy} loss function. The dropout rate is set to 0.3.}
\label{table:1D_CNN_architecture}
\end{table}

We compile the model with the \texttt{Adam} optimizer along with the default value for the learning rate and the \textit{sparse categorical crossentropy} loss function.
The model is pre-trained for 10 epochs with a batch size of 32 samples and then trained for further 60 epochs.
We save the model weights corresponding to the highest validation dataset accuracy.
The training is performed on the Case Western Reserve University HPC Pioneer facilities using the GPU cores powered by the Nvidia Tesla V100-SXM2-32GB V100 Graphics Accelerator Card.

\subsection{2D convolutional neural networks}
\label{appendix:2D_CNN_ResNet-50}

In order to directly capture the correlations between the different multipoles, we train a 2D convolutional neural network directly on the correlation data, i.e., $\mathcal{C}_{\ell m \ell^{\prime} m^{\prime}}$, rather than on the $a_{\ell m}$ for each realization.
This imposes tighter memory constraints, which limit us to $\ell_{\rm max} = 20$ for each realization.
The data is then prepared by extracting the $a_{\ell m}$ corresponding to $\ell \in [2, 20]$ for each realization and then calculating the corresponding $\mathcal{C}_{\ell m \ell^{\prime} m^{\prime}} = a_{\ell m} a_{\ell^{\prime} m^{\prime}}^{*}$.
We disregard the first two multipole values as we want our data to be comparable with the analytic correlation matrices (\cref{fig:cm_analytic_plots}) that are calculated without taking into account $\ell = 0$ and $\ell = 1$ data.
Similarly, this is done to correctly rescale the data by $\left(C_{\ell}^{\Lambda \mathrm{CDM}} C_{\ell^{\prime}}^{\Lambda \mathrm{CDM}}\right)^{1/2}$, which is calculated without taking the first two multipole values into account.
Since our used 2D CNN cannot natively deal with complex values, we split the data into components, which are stacked into 3 channels corresponding to the absolute values, the real part, and the imaginary part, i.e., $\mathcal{C}_{\ell m \ell^{\prime} m^{\prime}}^{i} = [\mathrm{Abs}(\mathcal{C}_{\ell m \ell^{\prime} m^{\prime}}^{i}), \mathrm{Re}(\mathcal{C}_{\ell m \ell^{\prime} m^{\prime}}^{i}), \mathrm{Im}(\mathcal{C}_{\ell m \ell^{\prime} m^{\prime}}^{i})]$.

\begin{figure}
  \centering
  \includegraphics[width=0.99\textwidth]{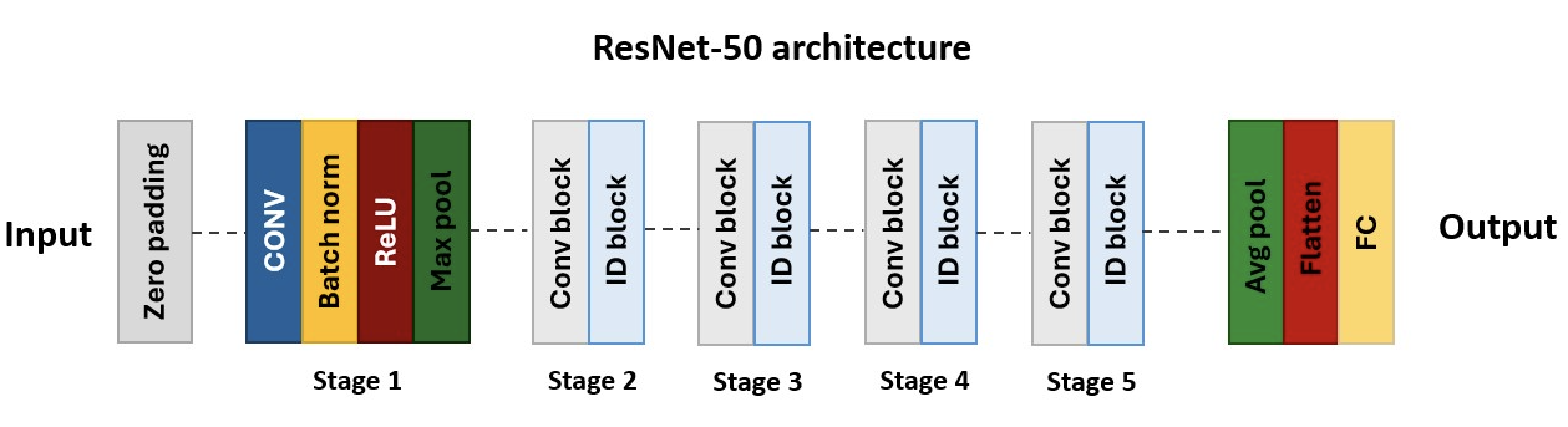}
  \caption{Main components of the \texttt{ResNet-50} architecture. The convolutional blocks refer to a series of convolutional layers followed by batch normalization and \texttt{ReLU} activation. The final convolutional layer in the convolutional block is generally appended with a shortcut/skip connection. The identity blocks consist of series of convolutional layers with $1 \times 1$ and $3 \times 3$ kernels followed by batch normalization and \texttt{ReLU} activations. In addition, the identity blocks are appended by shortcut/skip connections and an addition operation, which combines the original input with the learnt features from the convolutional layers in the block. The final block that is appended to the base \texttt{ResNet-50} architecture (``FC'') contains two \textit{Dense} layers with 64 and 32 neurons followed by dropout layers with dropout rates of 0.5. Full technical details on the \texttt{ResNet} architecture are provided in \rcite{Resnet-50-2015}. Reproduced from \rcite{ResNet-50-diagram}. CC BY 4.0. This ResNet50 image has been obtained by the authors from the Wikimedia website where it was made available by Gorlapraveen123 under a CC BY-SA 4.0 license. It is included withing this article on that basis. It is attributed to Gorlapraveen123.} 
  \label{fig:resnet50-architecture}
\end{figure}

For this particular dataset, we choose to work with the \texttt{ResNet-50} architecture \cite{Resnet-50-2015, Bello2021}.
\texttt{ResNet-50} is a deep neural network architecture consisting of a total of 50 layers, originally designed to tackle the vanishing gradient problem.
As a core feature, it uses the so-called residual blocks, each of which contains skip connections introduced to avoid the gradient values getting reduced to zero, which effectively halts the training procedure.
The bulk of the \texttt{ResNet-50} architecture consists of convolutional blocks followed by layers that perform identity mapping (where the input to a layer is directly added to the output).
These identity mapping layers, or identity blocks, are a type of residual blocks that combine multiple convolutional layers, followed by batch normalization layers with the \texttt{ReLU} activation.
The mentioned features along with the different filter sizes and pooling operations in the different layers of the architecture lead to \texttt{ResNet-50} being a promising choice for extracting features that are correlated at different scales in the training data (e.g., like the \textit{checkerboard} pattern observed in \cref{fig:cm_analytic_plots}).
While originally developed for image classification, the \texttt{ResNet}-based architectures have been used when working with much more abstract data, e.g., weak lensing convergence and mass maps \cite{Hong2021}, strong gravitational lensing data \cite{Liu2021}, and astronomical target classification \cite{Jia2020}.
The key components of the \texttt{ResNet-50} architecture are summarized in \cref{fig:resnet50-architecture}.

\texttt{ResNet-50} architecture is usually trained in one of the two ways: via transfer learning, or by fully retraining the model weights.
As an initial approach, we test the transfer learning approach by freezing the bulk of the model weights to their optimal values (i.e., model weights deduced by training a \texttt{Keras} \texttt{ResNet-50} model on the \texttt{ImageNet} dataset \cite{Russakovsky2014}) and then appending two fully-connected layers (followed by batch normalization and droupout layers), which are meant to fine-tune the pre-trained model weights.
This results in a sub-optimal performance, hence we choose the second approach, which is to fully re-train the \texttt{ResNet-50} model weights. We train the model on a total of 40,000 randomly rotated realizations, as well as on the same number of non-rotated data samples.
Following a procedure similar to the 1D CNN model, we pre-train the \texttt{ResNet-50} model for 30 epochs followed by further 40 epochs with a batch size of 32 data samples.
We save the best model during the second stage of the training procedure based on the maximum value of the validation dataset accuracy. 
The model is trained by using the Case Western Reserve University HPC Pioneer GPU nodes equipped with a Nvidia Tesla V100-SXM2-32GB V100 Graphics Accelerator Card.
Additional tests and optimization are done on the Pitzer Cluster provided by the Ohio Supercomputer Center using the high memory nodes with no GPU support \cite{OhioSupercomputerCenter1987}.

\subsection{Complex convolutional neural networks}
\label{appendix:CVNN}

Complex-valued convolutional neural networks allow us to use complex datasets as an input and classify them while maintaining the power of convolutional layers, i.e., studying the correlations between different multipoles.
Here, instead of feeding 3 channels of the covariance matrix as described in \cref{appendix:2D_CNN_ResNet-50}, we can directly use the complex matrix as the input.
As before, to satisfy the memory constraints, the dataset is formed by calculating $\mathcal{C}_{\ell m \ell^{\prime} m^{\prime}}$ from the $a_{\ell m}$ corresponding to $\ell \in [2, 20 ]$ for each realization. 

As illustrated in \cref{fig:cm_samples_norm}, cosmic variance introduces noise in the covariance matrix, smoothing away the expected patterns observed in the analytic matrices (\cref{fig:cm_analytic_plots}).
For this reason we choose to work with a particular CVNN architecture inspired by the one used in \rcite{Ribli:2019wtw}, which has been shown to be effective when working with noisy data.
The network is formed by 15 complex convolutional layers, organized in 5 sets followed by a complex average pooling layer (two sets of 2, two sets of 3, and one set of 5 convolutional layers). 
The output is flattened and then passed through to dense layers that are responsible for outputting the classification results (see \cref{table:CVNN_architecture}).
We train the model on a total of 40,000 randomly rotated realizations, as well as the same number of non-rotated data samples, using 80\% for training, 10\% for validating, and 10\% for testing. The network is pretrained for 40 epochs followed by further 80 with a batch size of 32 data samples. 
We save the best model during the second stage of the training procedure based on the maximum value of the validation dataset accuracy.
The model is trained by using the Nvidia Tesla V100-SXM2-32GB V100 Graphics Accelerator Card provided by the Pioneer HPC facilities at Case Western Reserve University.

\begin{table}[t!]
\centering
\begin{tabular}{lccc}
\hline \hline & \textbf{Activation} & \textbf{Output shape} & \textbf{Parameters} \\
\hline \hline Input map & - & (207,207,1,32) & - \\
\hline \texttt{ComplexConv2D} & \texttt{cart\_relu} & (None, 205, 205, 32) & 640 \\
 \texttt{ComplexConv2D}  & \texttt{cart\_relu}  & (None, 203, 203, 32)  & $18 \mathrm{~K}$ \\
\texttt{ComplexAvgPooling2D} & - & (None, 101, 101, 32)  & - \\
\hline \texttt{ComplexConv2D} & \texttt{cart\_relu} & (None, 99, 99, 64)  & $37 \mathrm{~K}$ \\ 
\texttt{ComplexConv2D} & \texttt{cart\_relu} & (None, 98, 98, 64)  & $33 \mathrm{~K}$ \\
\texttt{ComplexAvgPooling2D} & - & (None, 49, 49, 64)  & - \\ \hline
\texttt{ComplexConv2D} & \texttt{cart\_relu} & (None, 47, 47, 128)  & $148 \mathrm{~K}$ \\ 
\texttt{ComplexConv2D} & \texttt{cart\_relu} & (None, 47, 47, 64)  & $17 \mathrm{~K}$ \\ 
\texttt{ComplexConv2D} & \texttt{cart\_relu} & (None, 45, 45, 128)  & $148 \mathrm{K}$ \\ 
\texttt{ComplexAvgPooling2D} & - & (None, 22, 22, 128) &  - \\\hline
\texttt{ComplexConv2D} & \texttt{cart\_relu} & (None, 20, 20, 256)  & $590 \mathrm{~K}$ \\ 
\texttt{ComplexConv2D} & \texttt{cart\_relu} & (None, 20, 20, 128)  & $66 \mathrm{~K}$ \\ 
\texttt{ComplexConv2D} & \texttt{cart\_relu} & (None, 18, 18, 256)  & $590 \mathrm{K}$ \\ 
\texttt{ComplexAvgPooling2D} & - & (None, 9, 9, 256) &  - \\ \hline
\texttt{ComplexConv2D} & \texttt{cart\_relu} & (None, 7, 7, 512)  & $2.4 \mathrm{~M}$ \\ 
\texttt{ComplexConv2D} & \texttt{cart\_relu} & (None, 7, 7, 256)  & $263 \mathrm{~K}$ \\ 
\texttt{ComplexConv2D} & \texttt{cart\_relu} & (None, 5, 5, 512)  & $2.4 \mathrm{~M}$ \\ 
\texttt{ComplexConv2D} & \texttt{cart\_relu} & (None, 5, 5, 256)  & $263 \mathrm{~K}$ \\ 
\texttt{ComplexConv2D} & \texttt{cart\_relu} & (None, 3, 3, 512)  & $2.4 \mathrm{~M}$ \\ 
\texttt{ComplexAvgPooling2D} & - & (None, 1, 1, 512) &  - \\ \hline 

\texttt{ComplexFlatten} & - & (None, 512) & - \\ 
\texttt{ComplexDense} & \texttt{cart\_relu} & (None, 64) & $66 \mathrm{~K}$ \\ 

 \texttt{ComplexDense} & \texttt{softmax\_real\_with\_abs} & 4 & $520$ \\
\hline \hline \multicolumn{3}{l}{ \textbf{Total trainable parameters: }} & $\mathbf{590  ~K}$ \\
\hline \hline
\end{tabular}
\caption{Summary of the CVNN architecture used for classification of complex $\mathcal{C}_{\ell m \ell^{\prime} m^{\prime}}$ data. The used complex activation functions and the complex layers are explained in full detail in \rcite{CVNN_docs}. The architecture is based on the one described in \rcite{Ribli:2019wtw}.  }
\label{table:CVNN_architecture}

\end{table}

\section{Generating realizations with $L \gtrsim L_{\rm LSS}$ via Cholesky decomposition}
\label{appendix:large_L_results}

In order to test the effectiveness of the algorithms outlined in this work for classifying large realizations of the $E_{1}$ topology, a large dataset is required.
A key challenge for obtaining large datasets for realizations with $L \gtrsim L_{\rm LSS}$ by numerically evaluating \eqref{eq:a_lm_realizations} is that of memory constraints. Therefore, we present here a different method for generating large $E_{1}$ realizations.
Specifically, we follow the numerical approach described in \rcite{COMPACT:2023_paper2A} to evaluate the covariance matrix for the considered topology class. The covariance matrix can then be used to generate novel $a_{\ell m}$ realizations via Cholesky decomposition (see \rcite{watkins2004, Mukherjee2014, COMPACT:2023_paper2A}). 
The main advantage of this method when compared to numerically evaluating \eqref{eq:a_lm_realizations} is that we can generate realizations significantly faster, with the caveat that the range of multipoles that we consider in our dataset, $\ell \in [2, \ell_{max}]$, is limited to $\ell_{max} = 10-30$. In other words, the Cholesky decomposition method allows generating realizations with significantly higher $L$ values at a cost of having significantly smaller $\ell_{\rm max}$.

\begin{figure}[!h]
    \centering
    \begin{subfigure}{0.49\textwidth}
        \centering
        \includegraphics[width=1.1\textwidth]{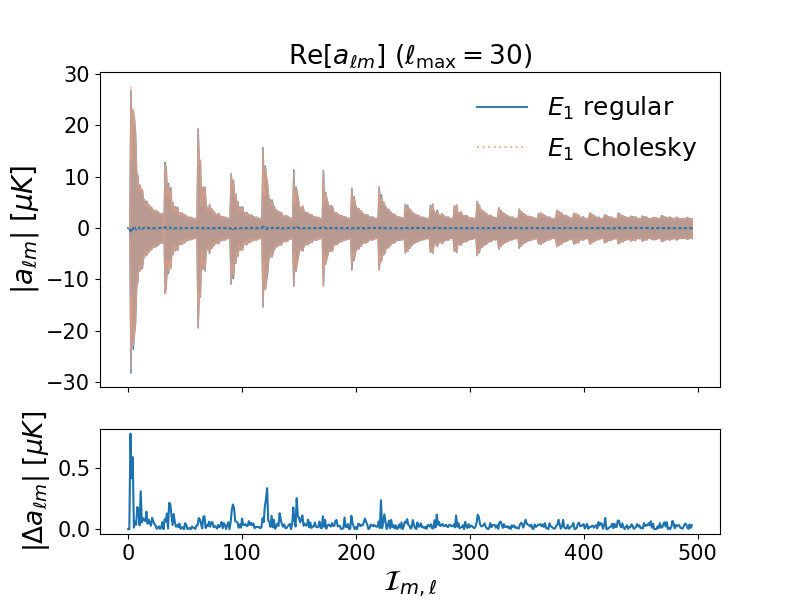}
        \label{fig:cholesky_vs_regular_1}
    \end{subfigure}
    \hfill
    \begin{subfigure}{0.49\textwidth}
        \centering
        \includegraphics[width=1.1\textwidth]{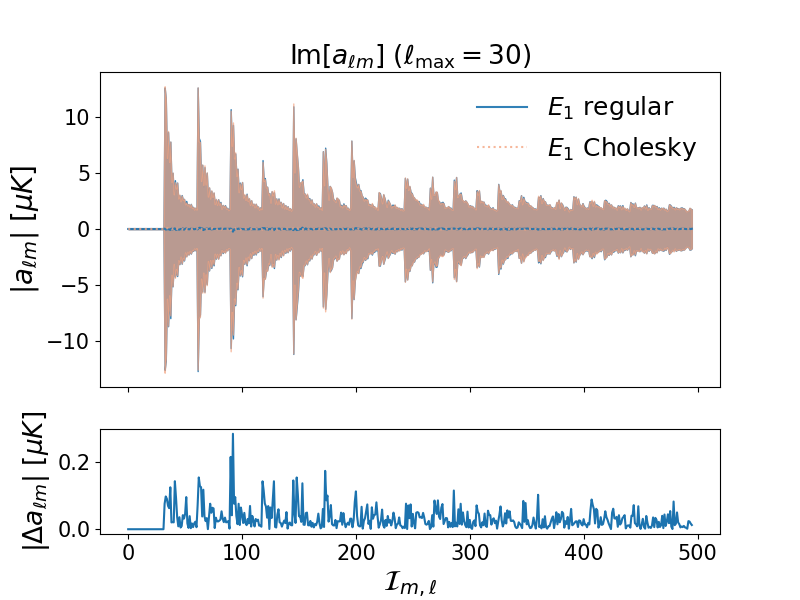}
        \label{fig:cholesky_vs_regular_2}
    \end{subfigure}
    \caption{A comparison of the real (left) and imaginary (right) values for an ensemble of 10,000 realizations generated via Cholesky decomposition and by numerically evaluating \eqref{eq:a_lm_realizations}. The solid and the dashed lines show the mean values for each $m$ and $\ell$, while the color bands correspond to the standard deviations. The residuals, i.e., the absolute values of the differences between the two mean value lines are shown as $\Delta a_{\ell m}$.}
    \label{fig:cholesky_vs_regular_re_im}
\end{figure}

\begin{figure}[!h]
  \centering
  \includegraphics[width=0.65\textwidth]{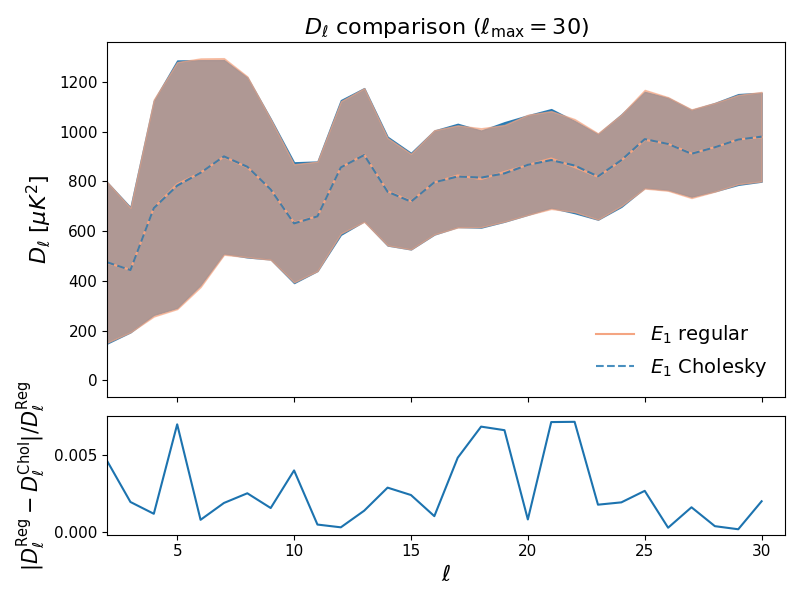}
  \caption{A comparison between the rescaled power spectra of the 10,000 realizations shown in \cref{fig:cholesky_vs_regular_re_im}. Note that $D_{\ell} \equiv \ell (\ell + 1) C_{\ell}/2 \pi$. The solid and the dashed lines show the mean values for each $\ell$, while the color bands correspond to the standard deviations. The residual ratio is shown in the bottom section of the figure. } 
  \label{fig:cholesky_vs_regular_cl}
\end{figure}

Since we are generating this particular set of realizations using a different method, a natural question is how similar are these realizations when compared to those obtained by numerically evaluating \eqref{eq:a_lm_realizations}?
To investigate this, we generate two sets of $E_{1}$ topology realizations with the same $\ell_{max}$ and $L$ values and perform a thorough comparison.
Namely, we generate a set of 100,000 realizations using the two methods, and then calculate the mean and the standard deviation for the real and imaginary values for each $\ell$ and $m$.
Similarly, we compute the mean and the standard deviation values for the power spectrum.
The obtained ensemble level quantities show only minor differences in the power spectrum and almost no discernible difference between the ensemble values of the $\mathrm{Re}[ a_{\ell m}]$ and $\mathrm{Im}[ a_{\ell m}]$ components.
The outlined comparison is illustrated in \cref{fig:cholesky_vs_regular_re_im,fig:cholesky_vs_regular_cl}.
Finally, to perform a comparison at an individual realization level, we train the random forest and 1D CNN classifiers to distinguish $a_{\ell m}$ realizations generated using the two distinct methods.
The obtained results show that neither random forests nor the 1D CNN is capable of distinguishing the two sets of realizations (i.e., the classification accuracy is $\sim 50 \%$).

\bibliographystyle{utphys}
\bibliography{topology,additional}

\end{document}